\pdfoutput=1

\documentclass[11pt,twoside,a4paper,cmspaper,final,collab]{cms-tdr}
\def\svnVersion{f524aa7-D}\def\svnDate{2020/05/07}\def\cmsCernNoTag{CERN-EP-2019-222}\def\cmsCernDate{\today}\def\cmsMessage{"Published in the Journal of High Energy Physics as \href{http://dx.doi.org/10.1007/JHEP05(2020)033}{\doi{10.1007/JHEP05(2020)033}.}"}

\begin{document}\cmsNoteHeader{EXO-19-012}

\hyphenation{had-ron-i-za-tion}
\hyphenation{cal-or-i-me-ter}
\hyphenation{de-vices}
\newcommand{\mjj}{\ensuremath{m_{\mathrm{jj}}}\xspace}
\newcommand{\detajj}{\ensuremath{\abs{\Delta\eta}}\xspace}
\newcommand{\Qstar}{\ensuremath{\PQq^*}\xspace}
\newcommand{\RunLumi}{137\fbinv}
\newcommand{\lumiUncert}{2.3\%\xspace}
\newcommand{\jecUncert}{2\%\xspace}
\newcommand{\RECOminMjjCut }{1.5\TeV}
\newcommand{\RECOminMjjCutControl}{2.4\TeV}
\newcommand{\mDM}{\ensuremath{m_{\text{DM}}}\xspace}
\newcommand{\gDM}{\ensuremath{g_{\text{DM}}}\xspace}
\newcommand{\gq}{\ensuremath{g_\PQq}\xspace}
\newcommand{\mMed}{\ensuremath{M_{\text{Med}}}\xspace}
\newcommand{\CRhigh}{\ensuremath{\mathrm{CR}_{\text{high}}}\xspace}
\newcommand{\CRmiddle}{\ensuremath{\mathrm{CR}_{\text{middle}}}\xspace}
\newcommand{\Rext}{\ensuremath{R}\xspace}
\newcommand{\Rextaux}{\ensuremath{R_{\text{aux}}}\xspace}

\cmsNoteHeader{EXO-19-012}
\title{Search for high mass dijet resonances with a new background prediction method in proton-proton collisions at $\sqrt{s}=13\TeV$}
\date{\today}

\abstract{A search for narrow and broad resonances with masses greater than 1.8\TeV decaying to a pair of jets is presented. The search uses proton-proton collision data at $\sqrt{s}=13\TeV$  collected at the LHC, corresponding to an integrated luminosity of \RunLumi. The background arising  from standard model processes is predicted with the fit method used in previous publications and with a new method. The dijet invariant mass spectrum is well described by both data-driven methods, and no significant evidence for the production of new particles is observed. Model independent upper limits are reported on  the production cross sections of narrow resonances, and broad resonances with widths up to 55\% of the resonance mass. Limits are presented on the masses of narrow resonances from various models: string resonances, scalar diquarks, axigluons, colorons, excited quarks, color-octet scalars, \PWpr and \PZpr bosons, Randall--Sundrum gravitons, and dark matter mediators. The limits on narrow resonances are improved by 200 to 800\GeV relative to those reported in previous CMS dijet resonance searches. The limits on dark matter mediators are presented as a function of the resonance mass and width, and  on the associated coupling strength as a function of the mediator mass. These limits exclude at 95\% confidence level a dark matter mediator with a mass of 1.8\TeV and width  1\%  of its mass or higher, up to one with a mass of 4.8\TeV and a width 45\% of its mass or higher.}

\hypersetup{%
pdfauthor={CMS Collaboration},%
pdftitle={A search for massive dijet resonances in proton-proton collisions at sqrts=13 TeV with a new background prediction method},%
pdfsubject={CMS},%
pdfkeywords={CMS, physics, dijets, resonances}}

\maketitle

\section{Introduction}

New particles that decay to pairs of jets and appear as dijet resonances arise in a variety of models.
String  resonances~\cite{Anchordoqui:2008di,Cullen:2000ef}  originate from the Regge excitations of quarks and gluons. Scalar diquarks~\cite{ref_diquark}  are predicted by a grand
unified theory based on the $E_6$ gauge symmetry group. Mass-degenerate excited quarks
(\Qstar)~\cite{ref_qstar,Baur:1989kv} appear in quark compositeness models. Axigluons and colorons, axial-vector and vector particles, are expected in the chiral
color~\cite{ref_axi,Chivukula:2013xla} and the flavor-universal coloron~\cite{ref_coloron,Chivukula:2013xla} models,
respectively. Color-octet scalars~\cite{Han:2010rf} appear in dynamical electroweak (EW) symmetry breaking models,
such as technicolor. New gauge bosons (\PWpr and \PZpr) can exist with standard model
(SM) like or leptophobic couplings~\cite{ref_gauge}. Randall--Sundrum (RS) gravitons are
predicted in the RS model of extra
dimensions~\cite{ref_rsg}. Dark matter (DM) mediators arise from an interaction
between quarks and DM~\cite{Chala:2015ama,Abercrombie:2015wmb,Abdallah:2015ter,Boveia:2016mrp}.
The natural width, $\Gamma$, of a new particle increases with its coupling strength to other states, and may vary from narrow to broad, as defined in comparison to
the experimental resolution.

This paper describes a model-independent search for a narrow or broad \textit{s}-channel dijet resonance with a mass
above 1.8\TeV, in proton-proton ($\Pp\Pp$) collisions at $\sqrt{s}=13\TeV$. This search uses data corresponding to an integrated luminosity of \RunLumi collected in
2016--2018 with
the CMS detector at the LHC. Similar searches  have been published previously by the ATLAS
and CMS Collaborations at $\sqrt{s}=13$\TeV~\cite{Aad:2019hjw,Sirunyan:2018xlo,Aaboud:2017yvp,Sirunyan:2016iap,Khachatryan:2015dcf,ATLAS:2015nsi},
8\TeV~\cite{Khachatryan:2016ecr,Khachatryan:2015sja,Aad:2014aqa,Chatrchyan:2013qhXX},
and 7\TeV~\cite{CMS:2012yf,Aad201237,ATLAS:2012pu,Chatrchyan2011123,Aad:2011aj,Khachatryan:2010jd,ATLAS2010} using
strategies reviewed in Ref.~\cite{Harris:2011bh}.  
Results of the search are interpreted using as benchmarks the models described above.  As no excess above the SM was observed, we set limits on the 
production cross sections of new particles decaying to the parton pairs $\PQq\PQq$ (or
$\PQq\PAQq$), $\Pq\Pg$, and $\Pg\Pg$. We then use these limits to constrain the benchmark  models, with the same choices of parameters as those that were used 
in the most recent CMS search~\cite{Sirunyan:2018xlo}, which used data corresponding to an integrated 
luminosity of 36\fbinv.
In the color-octet scalar 
model, the squared anomalous coupling
$k_\mathrm{s}^2=1/2$~\cite{Chivukula:2014pma} is used. For the RS graviton model, the value of the dimensionless coupling $k/\overline{M}_\text{Pl}$ is chosen to be
0.1, where $k$ is the curvature scale in the 5-dimensional anti de Sitter space and $\overline{M}_\text{Pl}$ is the
reduced Planck scale defined as $M_\text{Pl}/\sqrt{8\pi}$. For the DM mediator, we follow the recommendations of
Ref.~\cite{Boveia:2016mrp} on model choice and coupling values. We use a  simplified model~\cite{Abdallah:2015ter} of a spin-1 mediator decaying
only to quark-antiquark ($\PQq\PAQq$) and DM particle pairs, with an unknown mass \mDM, and with a universal quark coupling $\gq = 0.25$ and a DM coupling
$\gDM=1.0$.

Similar to past searches, and for dijet mass ($\mjj$) greater than 1.5\TeV, the main background from quantum
chromodynamics (QCD) multijet production is predicted by fitting the $\mjj$ distribution with an empirical  functional form. For $\mjj>2.4$\TeV,
a new  data-driven method is introduced, which predicts the  background from a control region where the pseudorapidity separation of the two jets, $\detajj$, is large.
This new "ratio  method" yields smaller systematic uncertainties when performed in the same dijet mass range  as the "fit method", and the sensitivity for broad resonance
searches is improved by up to a factor of two depending on the resonance
width and mass. In addition, the total integrated luminosity for this search is roughly
a factor of four larger than that used by the previous CMS search~\cite{Sirunyan:2018xlo},
so the sensitivity of both narrow and broad resonance searches has also
increased by up to an additional factor of two.

\section{The CMS detector}
\label{sec:detector}

A detailed description of the CMS detector and its coordinate system, including definitions of the azimuthal angle $\phi$ and
pseudorapidity $\eta$, is given in Ref.~\cite{refCMS}.
The central feature of the CMS apparatus is a superconducting
solenoid of 6\unit{m} internal diameter providing an axial magnetic field of 3.8\unit{T}.
Within the solenoid volume are located the silicon pixel and strip tracker ($\abs{\eta}<2.4$), and the barrel and
endcap calorimeters ($\abs{\eta}<3.0$), where these latter detectors consist of a lead tungstate crystal electromagnetic
calorimeter and a brass and scintillator hadron calorimeter.
An iron and quartz-fiber hadron calorimeter is located in the forward region ($3.0<\abs{\eta}<5.0$),
outside the solenoid volume. The muon detection system covers $\abs{\eta}<2.4$ with up to four layers of gas-ionization chambers installed
outside the solenoid and embedded in the layers of the steel flux-return yoke.

\section{Jet reconstruction and event selection}
\label{sec:reco}

A particle-flow (PF) event algorithm aims to reconstruct and identify each individual particle in an event, with an optimized
combination of information from the various elements of the CMS detector~\cite{CMS-PRF-14-001}.
Particles are classified as muons, electrons, photons, charged hadrons, or neutral hadrons.  To reconstruct jets, the anti-\kt algorithm~\cite{Cacciari:2005hq,Cacciari:2008gp} is used with a
distance parameter of 0.4, as implemented in the \textsc{FastJet} package~\cite{Cacciari:2011ma}. At least one reconstructed vertex is required.
Charged PF candidates not originating from the primary vertex
are removed prior to the jet finding. The candidate vertex with the largest value of summed physics-object $\pt^2$, where \pt is the transverse momentum, is taken to be the primary $\Pp\Pp$
interaction vertex. The
physics objects are the jets, clustered
using the jet finding algorithm mentioned above, with the tracks assigned to candidate vertices as inputs, and the associated missing transverse momentum,
taken as the negative vector sum of the \pt of those jets.
For jets, an event-by-event correction based on jet
area~\cite{jetarea_fastjet_pu,Khachatryan:2016kdb}
is applied to the jet energy to remove the estimated contribution from additional collisions in
the same or adjacent bunch crossings (pileup).

Events are selected using a two-tier trigger system~\cite{Khachatryan:2016bia}. Events satisfying
loose jet requirements at the first-level (L1) trigger are examined by the high-level trigger (HLT) system. Single-jet
triggers that require a jet in the event to exceed a predefined \pt threshold are used. Triggers that require \HT to exceed a threshold, where \HT is the scalar sum of jet \pt for all jets in the event with $\pt>30$\GeV
and $\abs{\eta}<3.0$, are also used. The HLT requires: \HT$>1050$\GeV or at least one jet reconstructed with an increased distance parameter of 0.8 and $\pt>550$\GeV.

The jet momenta and energies are corrected using calibration factors obtained from simulation, test beam results, and $\Pp\Pp$ collision
data at $\sqrt{s}=13$\TeV. The methods described in Ref.~\cite{Khachatryan:2016kdb} are used and all \textit{in-situ}
calibrations are obtained from the current data. Jets are required to have $\pt>30$\GeV and $\abs{\eta}<2.5$.  The two jets with the largest
\pt are defined as the leading jets. Jet identification criteria are applied to remove spurious jets associated with the calorimeter noise as well as those
associated with muon and electron candidates that are either misreconstructed or isolated~\cite{CMS:2017wyc}.
For all jets, we require that the neutral hadron and
photon energies are less than 90\% of the total jet energy.
For jets within the fiducial tracker coverage, we additionally require the jet to have nonzero charged-hadron
energy, and electron and muon energies to be less than 90\% and 80\% of the
total jet energy respectively. An event is rejected if either of the two
leading jets fails these jet identification criteria.

Each of the two leading jets is formed into a ``wide jet'' using an algorithm introduced for previous CMS dijet searches in Ref.~\cite{Khachatryan:2015sja}.
This wide-jet algorithm, designed
for dijet resonance event reconstruction, reduces the sensitivity of the analysis to gluon radiation (g) from the
final-state partons.  The two leading jets are used as seeds and the four-vectors of all other jets, if within a distance defined as $\sqrt{\smash[b]{(\Delta\eta)^2 +(\Delta\phi)^2}}<1.1$, are added to the
nearest leading jet to obtain two wide jets, which then form
the dijet system. The dijet mass is then found as the invariant mass of the system of these two wide jets.
The wide-jet algorithm thereby collects hard-gluon radiation found near the leading two final-state partons, in order to improve the dijet mass
resolution.

The background from $t$-channel dijet events has the same angular distribution as Rutherford scattering,
approximately proportional to $1/[1-\tanh(\detajj/2)]^2$, which peaks at large values of $\detajj$, the pseudorapidity separation of the two jets.
The signal region (SR) is defined by requiring $\detajj<1.1$, which maximizes the search sensitivity for
isotropic decays of dijet resonances in the presence of QCD dijet background.
For the ratio method of estimating the background, two control regions (CRs) are defined from events within $1.1<\detajj<2.6$.
The primary control region, $\CRhigh$, which contains events that satisfy
$1.5<\detajj<2.6$, is used to predict the main QCD background in the SR.  The secondary control region, $\CRmiddle$,
which contains events that satisfy $1.1<\detajj<1.5$, is used to constrain theoretical  and experimental systematic uncertainties.
The $\CRhigh$ is defined such that it has four to five times more background events than the SR, and at the same time fewer
signal events by a factor of two. The SR is used to search for  the presence of resonances and to estimate the QCD background for the fit method.

Events with $\mjj>\RECOminMjjCut$ are selected offline, for which the $\detajj$ between the two jets is in the interval
$\detajj<2.6$,  where the dijet mass and $\detajj$ are  reconstructed  using wide jets.
For this selection the combined L1 trigger and HLT was found to be fully efficient, as measured using a sample acquired with an independent trigger requiring at
least one muon with $\pt>50$\GeV at the HLT. The $\detajj<1.1$ requirement makes the trigger efficiency increase sharply and plateau at a value of 100\% for
relatively low values of dijet mass. This is because the jet \pt threshold of the trigger at a fixed dijet mass is more easily satisfied at
low $\detajj$, as seen by the approximate relation $\mjj\approx 2\pt\cosh(\detajj/2)$. Hence, the trigger efficiency reaches 100\% in
the SR at a lower value of dijet mass (1.5\TeV) than in both CRs (2.4\TeV). Therefore the fit method is used for $\mjj>1.5$\TeV and the ratio method, which requires data from the CRs with
100\% trigger efficiency, is used for $\mjj>2.4$\TeV.

\section{Data and simulation comparison}

As the dominant background for this analysis is expected to be the QCD production of two
or more jets, the selected dijet data are compared with QCD predictions.
The predictions come from 270 million simulated events produced by the \PYTHIA8.205~\cite{Sjostrand:2014zea}
program with the CUETP8M1 tune~\cite{Khachatryan:2015pea,Skands:2014pea} using the parton distribution function (PDF) set NNPDF2.3LO~\cite{Ball:2012cx},
including a \GEANTfour-based \cite{refGEANT} simulation of the CMS detector. The data-over-simulation ratio of event yields is $0.94$.
This search uses the signal shapes of narrow and broad
resonances presented in
Ref.~\cite{Sirunyan:2018xlo}, which are also from a \PYTHIA simulation.

\begin{figure}[htbp]
  \centering
   \includegraphics[width=0.98\textwidth]{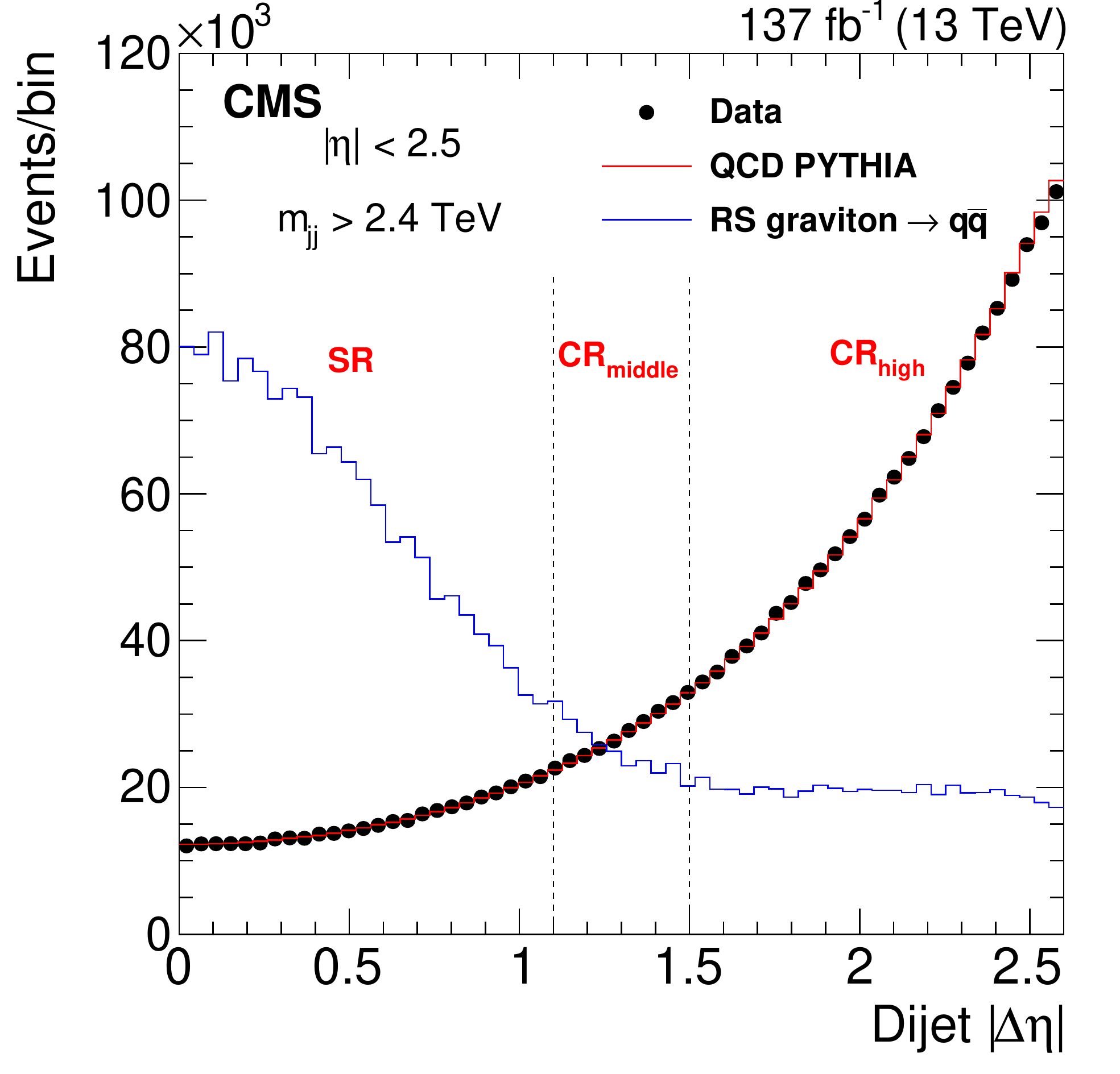}
   \caption{The pseudorapidity separation between the two wide jets
for the signal and control  regions. Data
(black points) are compared to QCD predictions from the \PYTHIA MC with detector simulation (red histogram) normalized to data. A signal from an
RS graviton decaying into a $\PQq\PAQq$ pair is also shown (blue histogram) normalized to data.}
    \label{figDeltaEta}
\end{figure}

The dijet $\detajj$ separation between the two wide jets is shown in Fig.~\ref{figDeltaEta}.
The data distribution shows that dijet production is dominated by
$t$-channel parton exchange, as predicted by QCD, with a production rate that increases with increasing $\detajj$.
By contrast, most \textit{s}-channel signals from dijet resonances decrease  with increasing $\detajj$, as the signal shown does.
Figure~\ref{figDeltaEta} shows
the division of the $\detajj$ distribution into the signal and control regions.

\begin{figure}[htbp]
  \centering
     \includegraphics[width=0.98\textwidth]{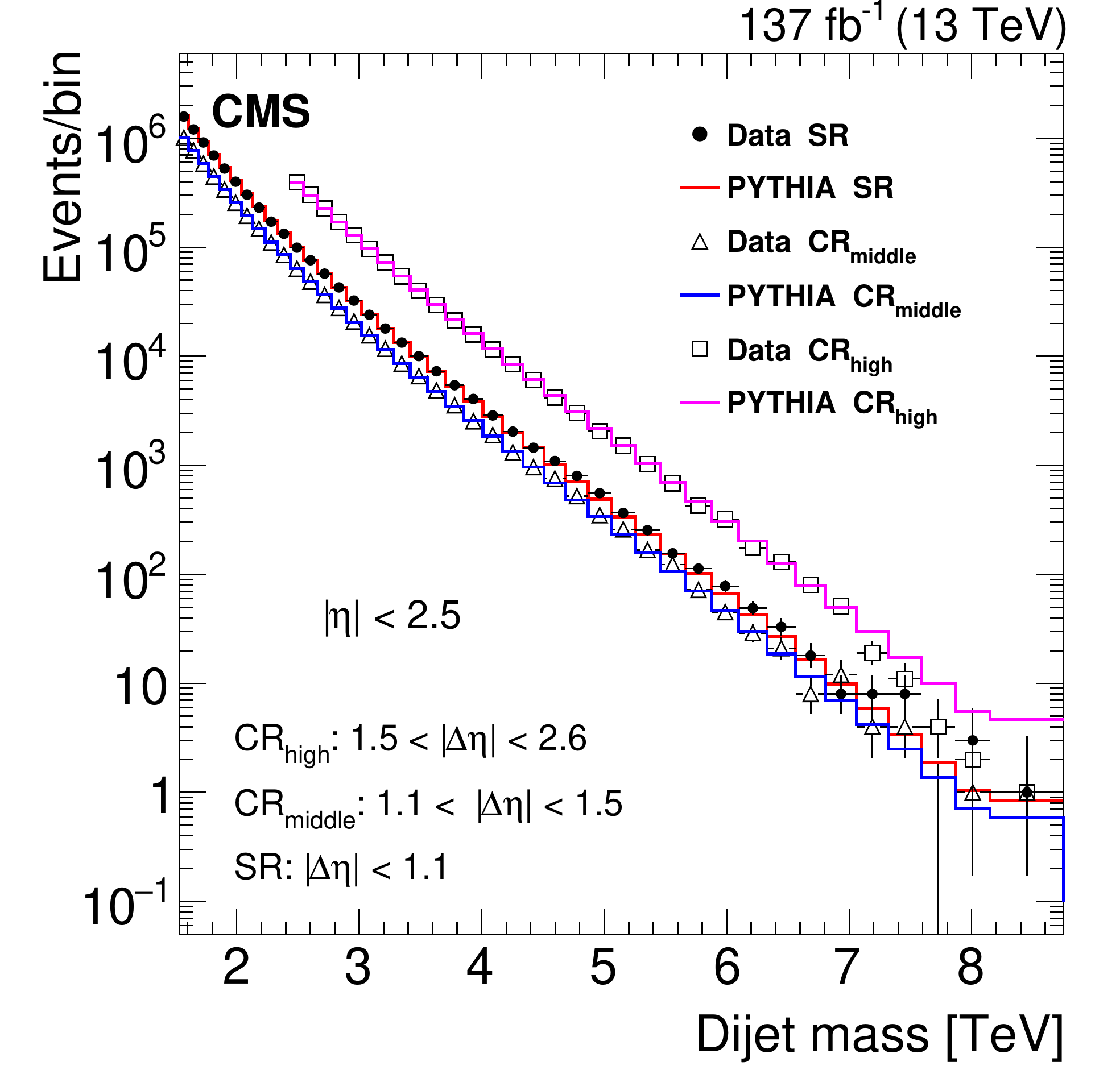} 
     \caption{The dijet mass spectra of the data and \PYTHIA simulation in the signal region at low $\abs{\Delta\eta}$ (black points and red histogram), 
     control region at middle $\abs{\Delta\eta}$ (triangles and blue
     histogram), and control region at high $\abs{\Delta\eta}$ (squares and magenta histogram). The simulation is normalized to data.}
    \label{figDijetMass}
\end{figure}

Figure~\ref{figDijetMass} shows, for both data and the QCD background,
the dijet mass spectra in the signal and control regions, which fall steeply and smoothly as a function of
dijet mass. The observed dijet mass distributions are compared to the QCD background prediction from \PYTHIA,
which simulates processes at leading order (LO).

\begin{figure}[htbp]
  \centering
   \includegraphics[width=0.98\textwidth]{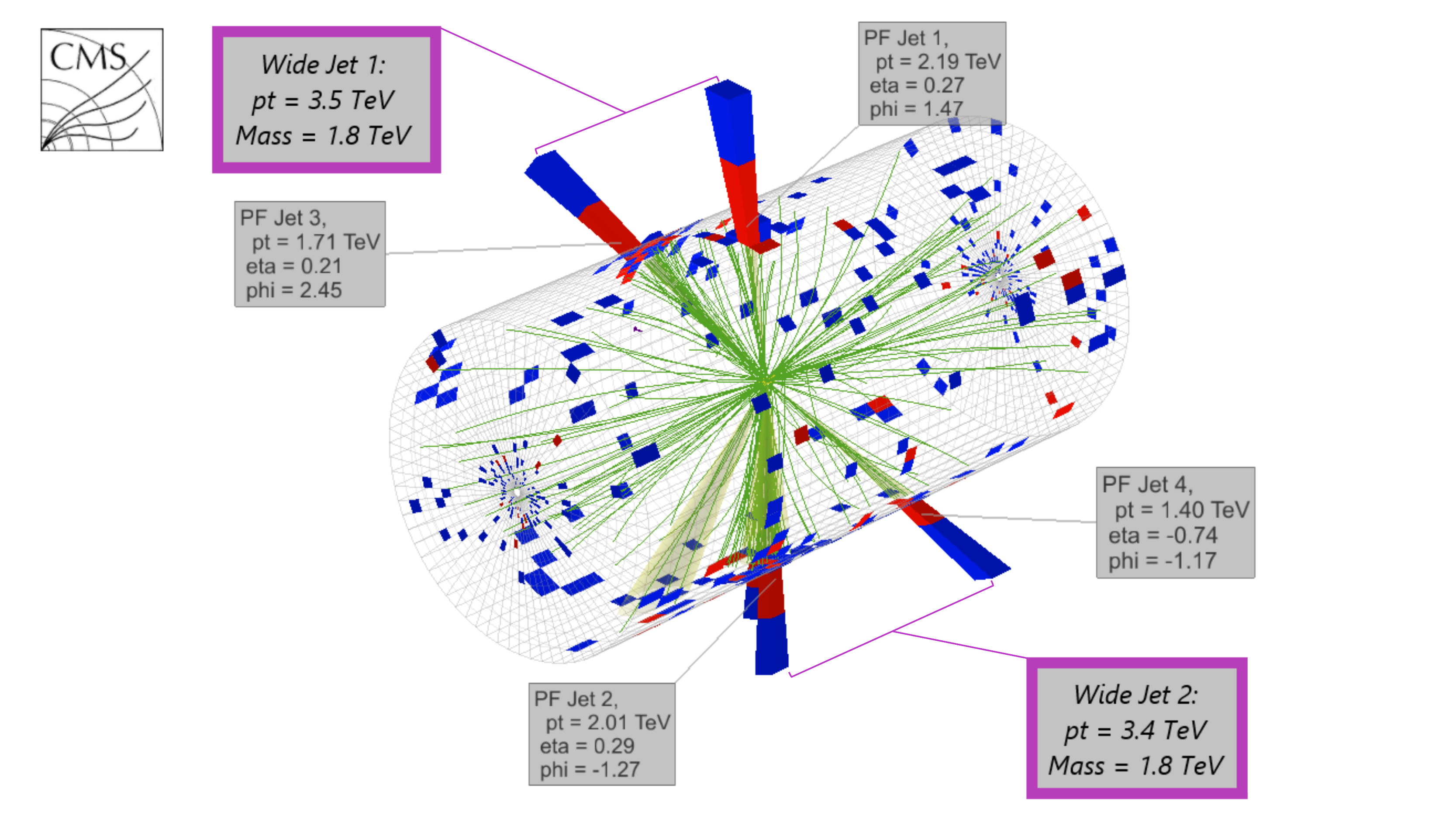}
   \caption{Three-dimensional display of the event with the second-highest dijet invariant mass of 8\TeV.  The display shows the energy deposited in the 
electromagnetic (red) and hadronic (blue)
  calorimeters and the reconstructed tracks of charged particles (green). The grouping of four
  observed jets into two wide jets (purple) is discussed in the text.}
    \label{eventsdisplay2}
\end{figure}

We inspect the characteristics of the 23 events with $\mjj>7$\TeV, to determine if they have the two-jet topology typical of the QCD background
and to check for the presence of detector and reconstruction pathologies,
and we find the one unusual event, shown in Fig.~\ref{eventsdisplay2}.
This event is the one with the second highest dijet mass, 8\TeV, and is unusual because it is composed of four jets, in two pairs, which are
combined into the two wide jets. It is also unusual because the wide jet mass, equal to the pair mass of the jets, has
the same value 1.8\TeV for each of the two wide jets.
The leading wide jet has a \pt of 3.5\TeV, and the other wide jet has a \pt of 3.4\TeV. The
wide jets are back-to-back in azimuthal angle ($\Delta\phi=3.1$) and nearby in pseudorapidity ($\detajj=0.4$).
Each one of the two wide jets is composed of two jets with cone size 0.4, with \pt, $\eta$, and $\phi$ values
as shown in Fig.~\ref{eventsdisplay2}.

The possibility that this event originates from a resonance decaying to a pair of dijet resonances has been
recently explored in a phenomenology paper~\cite{ref_dobrescu}.

\section{Background prediction methods}

\begin{figure}[htbp]
  \centering
  \includegraphics[width=0.98\textwidth]{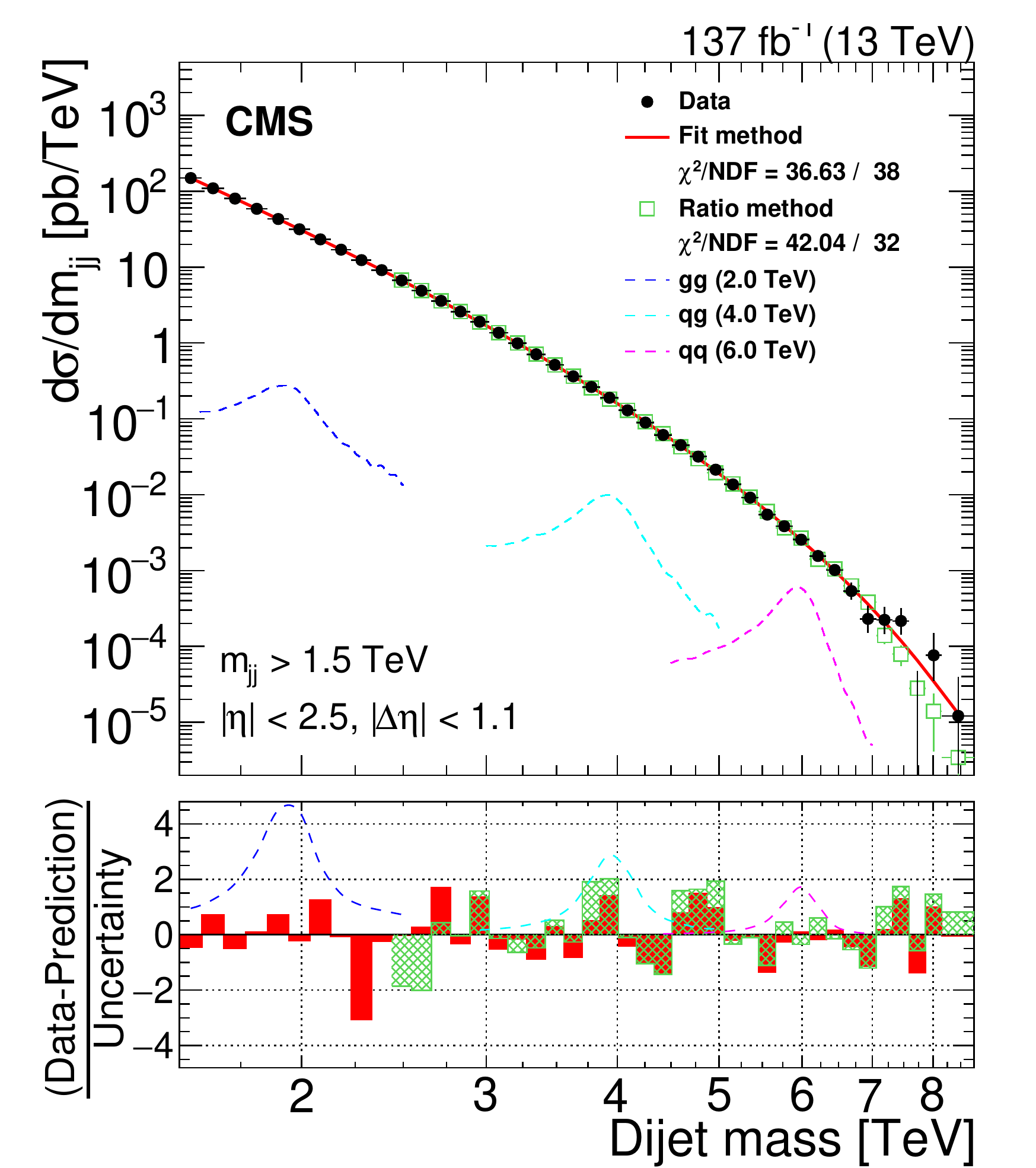}
  \caption{   
  Dijet mass spectrum in the signal region (points) compared to a fitted
  parameterization of the background (solid line) and the one obtained
  from the control region (green squares).
  The lower panel shows the difference between the data and the
  fitted parametrization (red, solid), and the data and the prediction obtained from the control region (green, hatched),
  divided by the statistical uncertainty in the data, which for the ratio method includes the statistical uncertainty in the data in the control region.
  Examples of predicted signals from narrow gluon-gluon, quark-gluon, and quark-quark resonances are shown (dashed coloured lines)
  with cross sections equal to the observed upper limits at 95\%~\CL.}
  \label{figDataAndFit}
\end{figure}

In the fit method, utilized here and in previous dijet resonance searches
~\cite{Sirunyan:2018xlo,Sirunyan:2016iap,Khachatryan:2016ecr,Khachatryan:2015dcf,Khachatryan:2010jd,Chatrchyan2011123,CMS:2012yf,Chatrchyan:2013qhXX,Khachatryan:2015sja,
ATLAS:2015nsi,ATLAS2010,Aad:2011aj,Aad201237,ATLAS:2012pu,Aad:2014aqa,refCDFrun2}, the main background in the SR coming from QCD is parametrized with
an empirical function of the form

\begin{equation}
\frac{{\rd}\sigma}{{\rd}\mjj} =
\frac{P_{0} (1 - x)^{P_{1}}}{x^{P_{2} + P_{3} \ln{(x)}}},
\label{eqBackgroundParam}
\end{equation}
where $x=\mjj/\sqrt{s}$, and $P_0$, $P_1$, $P_2$, and $P_3$ are four free parameters. The search for resonances proceeds with fitting the dijet mass 
distribution in the SR using this background parametrization and the signal template obtained from simulation, a procedure denoted as a signal plus 
background fit. In this fit, $P_0$, $P_1$, $P_2$, and $P_3$ are treated as freely floating nuisance parameters. In order to examine the compatibility
of the data with the background-only description, and the quality of the background prediction, a fit under only the background hypothesis, denoted
as a background-only fit, is also performed.
The chi-square per the number of degrees of freedom of the background-only fit is $\chi^2/\mathrm{NDF}=36.63/38$, as shown
in Fig.~\ref{figDataAndFit}.

\begin{figure}[htbp]
  \centering
     \includegraphics[width=0.48\textwidth]{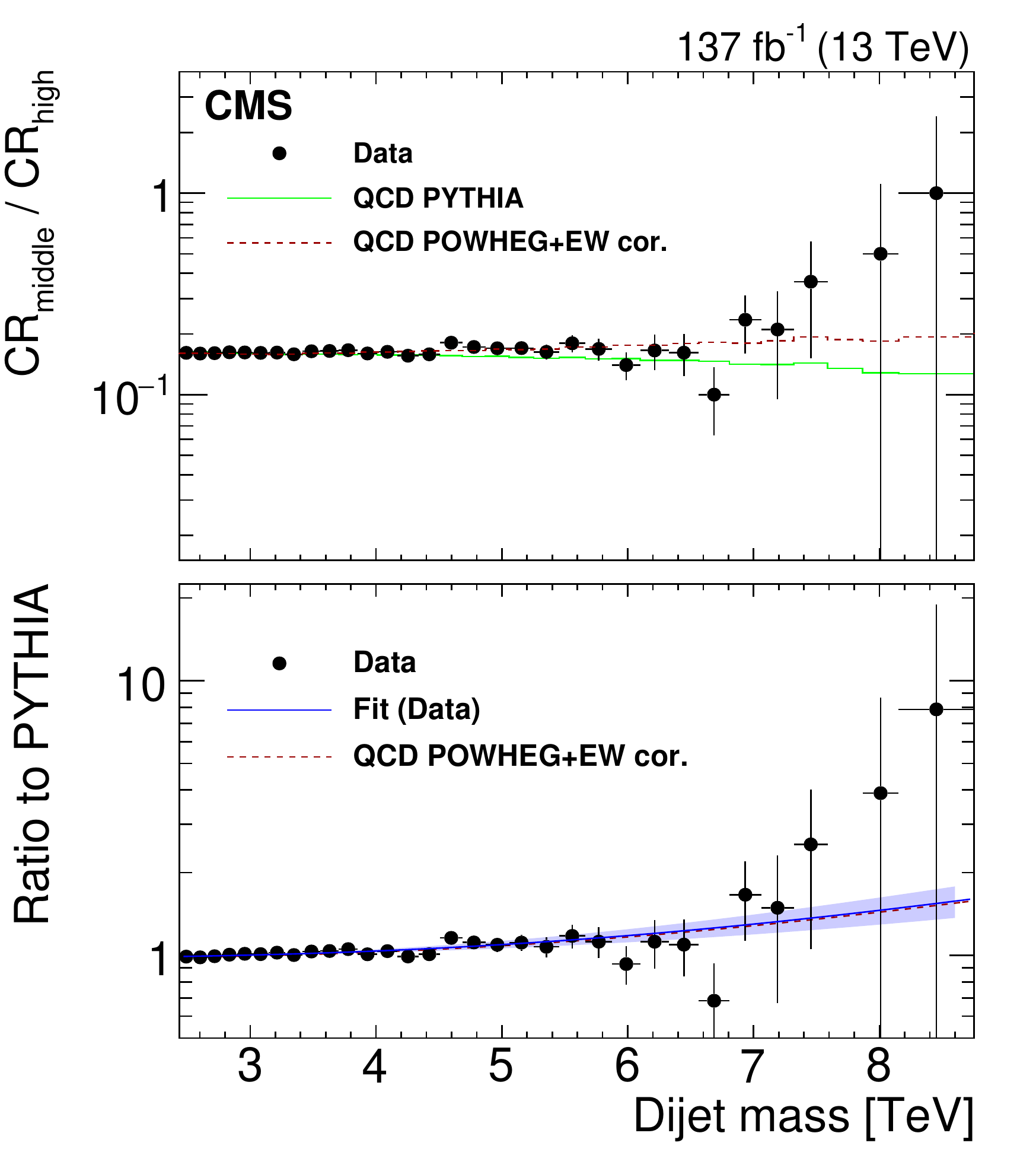}
     \includegraphics[width=0.48\textwidth]{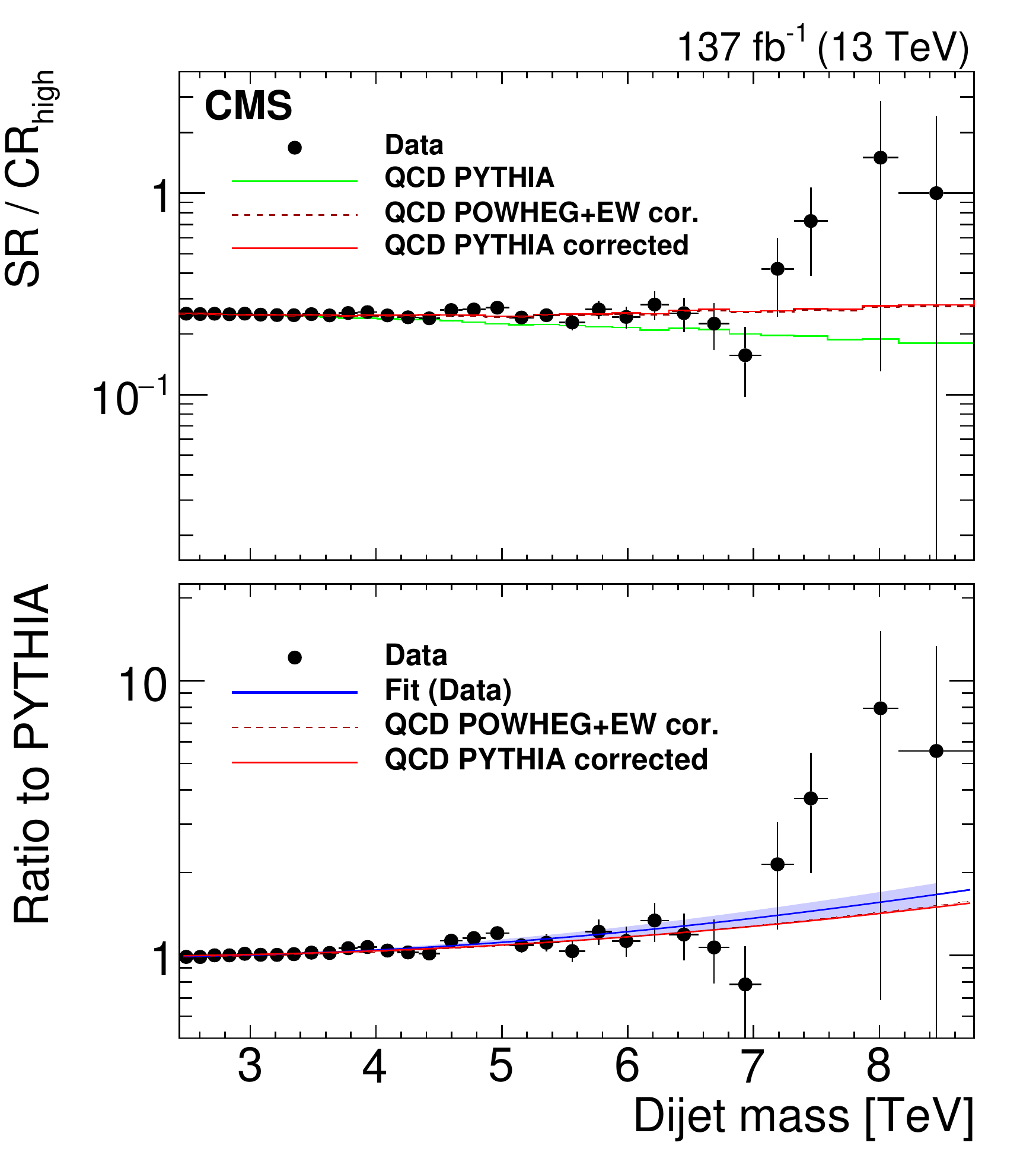}
     \caption{
      The ratio $\Rextaux$, the auxiliary transfer factor, calculated for data, \PYTHIA, and \POWHEG with
      electroweak corrections (left, upper panel). The double ratio of the same quantities in the upper left panel to
      $\Rextaux$ from \PYTHIA, along with the fit of the double ratio for data with the correction function (left, lower panel).
      The ratio $\Rext$,  the transfer factor, calculated for data,  \PYTHIA,
      \POWHEG with electroweak corrections, and corrected \PYTHIA (right, upper panel). The double ratio of
      the same quantities in the upper right panel to
      $\Rext$ from \PYTHIA, along with the fit of the double ratio for data with a correction function, and corrected
      \PYTHIA using $\CRmiddle$ (right, lower panel). The fits in the two lower panels agree with each other within their uncertainty at 95\%~\CL
      (shaded bands).}
    \label{figDataControlsRatios}
\end{figure}

The ratio method is a data-driven prediction of the QCD background in the SR, obtained by
multiplying the data in $\CRhigh$ by a mass-dependent transfer factor determined from the simulated angular
distribution of QCD dijet production. The transfer factor is the ratio, $R$, between the simulated dijet mass distribution of
background events in the SR and $\CRhigh$. The method makes use of the following definitions:
\begin{equation*}
\begin{aligned}
  N(i)^{\text{Prediction}}_{\mathrm{SR}} &= \Rext(\mjj/\sqrt{s}) N(i)^{\text{Data}}_{\CRhigh}, \\
 \Rext(\mjj/\sqrt{s}) &= C(\mjj/\sqrt{s}) N(i)^{\text{Sim.}}_{\mathrm{SR}} / N(i)^{\text{Sim.}}_{\CRhigh},
\end{aligned}
\end{equation*}
where $N(i)$ is the number of events in a given bin, $i$, of dijet mass and $C(\mjj/\sqrt{s})$ is a correction to the simulated
transfer factor. This correction is required because, as seen in the upper right panel of  Fig.~\ref{figDataControlsRatios},
differences are present between data and the simulation using \PYTHIA.
These are due to both theoretical and experimental effects.
The theoretical effects arise because the \PYTHIA simulation uses a QCD calculation at LO, and higher order QCD corrections have some
effect, and so do missing EW corrections.
Figure~\ref{figDataControlsRatios} shows, with a smaller sample of events, that a better agreement is obtained when these corrections are included, by generating events
at next-to-leading order (NLO) in QCD with \POWHEG v2.0~\cite{Nason:2004rx,Frixione:2007vw,Alioli:2010xd} and incorporating an estimate of EW effects ~\cite{Dittmaier:2012kx}.
Experimental effects include differences between data and simulation at higher jet pseudorapidities outside
the barrel calorimeter region ($\abs{\eta}>$1.3). The higher-order QCD and EW effects, and the differences between data and simulation
at higher jet pseudorapidity values, produce a similar effect on the shape of the transfer factor, affecting mainly the high dijet mass region.
We correct the simulated transfer factor to include these effects in a data-driven way, using the  second 
control region, $\CRmiddle$, which is a $\abs{\Delta\eta}$ sideband to the SR. This second control 
region contains dijet events with values of jet pseudorapidity very similar to those in the SR, 
and has a very small signal contamination. As such, the dijet mass distribution of this control region is very similar to that of the SR, and the 
differences between data and simulation in this control region are caused by similar theoretical and 
experimental effects as observed in the SR. Hence, this second CR allows  the definition of an auxiliary 
transfer factor, $\Rextaux$, shown in Eq.~(\ref{rext_aux}).

\begin{equation}
\Rextaux(i) =  N(i)_{\CRmiddle} / N(i)_{\CRhigh}.
\label{rext_aux}
\end{equation}

We then estimate the correction, $C$, to the main transfer factor, $\Rext$, by performing a fit to the data-over-simulation ratio of $\Rextaux$ 
(Eq.~(\ref{eqcor2})):

\begin{equation}
R_{\text{aux}}^{\text{Data}} / R_{\text{aux}}^{\text{Sim.}},
\label{eqcor2}
\end{equation}

with the correction parametrized using a two-parameter empirical function, shown in Eq.~(\ref{eqcor}).

\begin{equation}
C(\mjj/\sqrt{s}) = p_0 + p_1 (\mjj/\sqrt{s})^{3}.
\label{eqcor}
\end{equation}

The data to simulation ratios of the two transfer factors, $\Rextaux$ and $\Rext$, along with their  background-only fits, performed
separately in order to examine their compatibility, are shown in the lower panels of Fig.~\ref{figDataControlsRatios} and agree to within their uncertainty at 95\% 
confidence level (CL). Specifically, the values of the parameters and their statistical uncertainties from the background-only fits of the data to  
simulation ratios of $\Rextaux$ are $p_{0}=0.977\pm0.004$ and $p_{1}=2.07\pm0.33$, and of $\Rext$ are $p_{0}=0.972\pm0.004$ and $p_{1}=2.52\pm0.28$, and are entirely 
compatible. This agreement is expected given that the events in $\CRmiddle$ and SR have, by construction, very similar
jet $\eta$ and $\mjj$ distributions. 
Parameters $p_0 $ and $p_1$ are treated as free nuisance parameters  in the final signal plus background
simultaneous fit of the SR, $\CRmiddle$ and $\CRhigh$, taking the signal contamination in the control regions into account
as described in the next paragraph. The simultaneous background-only fit yields $p_{0}=0.973\pm0.003$ and $p_{1}=2.38\pm0.23$, consistent
with the  separate background-only fits shown in Fig.~\ref{figDataControlsRatios} (lower panels), and with smaller uncertainty.
The systematic uncertainty in the background, for both methods,  is automatically evaluated via profiling. This effectively
refits for the optimal values of the background parameters, allowing them to float freely, for each value of the resonance cross section.

The signal contamination in the CRs depends on the angular distribution of
the model. For the models considered in this search, the signal contamination is small compared to the
background. This is because we search for dijet resonances produced in the \textit{s}-channel annihilation of
two partons, while the QCD background is predominantly a $t$-channel process. We assume the signal has the same angular
distribution as a  vector resonance decaying to $\PQq\PAQq$ pairs. The signal contamination
is taken into account in the simultaneous fit. The change in extracted
signal is negligible if the angular distribution of any of our other benchmark models is chosen
instead. Our benchmark models include scalars coupled to $\PQq\PQq$ or $\Pg\Pg$ pairs, fermions coupled to $\Pq\Pg$ pairs, vectors coupled to
$\PQq\PAQq$ pairs, and tensors coupled to $\PQq\PAQq$ or $\Pg\Pg$ pairs.

Detailed signal injection tests are performed to investigate the potential bias in each background prediction method, and the bias is found to be 
negligible when either the fit method or the ratio method is employed. The signal injection tests are performed as follows: 
pseudo-data distributions are generated, varying the parameters of the 
background prediction and injecting a signal with a cross section equal to  i) zero, ii) the 95\%~\CL observed limit,  and iii) 
two times the 95\%~\CL observed limit. These distributions are created for several resonance masses and widths, spanning the entire 
range for which results are reported. Then, the same fitting procedure followed in the analysis of the actual  data is repeated for 
each pseudo-data  distribution, and the fitted signal  cross section, along with its 68\%~\CL standard deviation, is obtained. We 
examine the  distribution of the bias in units of standard deviations, namely the difference between the injected signal cross section 
and the  fitted signal cross section, divided by the standard deviation of the fit.  For all resonance masses, widths, and signal 
strengths  considered, the mean bias is less than one half a standard deviation, and in the vast majority of the cases it is well 
below this criterion.  In addition, pseudo-data distributions are  generated using different empirical functional forms than the ones 
used in the actual data fits, and the entire procedure is repeated, again yielding negligible biases.

The ratio method is an independent approach compared to the fit method, yielding consistent results.
The ratio method provides a background estimate that is derived primarily from control regions, while
the fit method uses only the signal region. The ratio method also provides a background estimate that is more accurate than the fit method. This is because
the ratio method fits the data with only two parameters, while the fit method requires four, and
because the estimate from the ratio method is additionally constrained by the control region $\CRmiddle$.
The advantages of this method, as opposed to the fit method, are the following: i) it provides a background estimate independent of the signal region,
which results in an independent and less biased value of the observed signal significance, ii) as the resonance width increases the ratio method has smaller
background uncertainty compared to the fit method, and hence higher sensitivity. Therefore, we estimate the background using the ratio method instead of
the fit method for $\mjj>\RECOminMjjCutControl$.

Figure~\ref{figDataAndFit} shows the dijet mass spectrum, defined as the observed number of events in each bin divided by the
integrated luminosity and the bin width. The bin widths depend on the dijet mass and are chosen to correspond to dijet mass resolution.
The bin edges were chosen  to be the same as those used by previous dijet resonances searches performed by the CMS Collaboration, as introduced in Ref.~\cite{Khachatryan:2010jd}.
Figure~\ref{figDataAndFit} also shows the background prediction from the fit method,
compared to all data, and the background prediction from the ratio method, compared to data with $\mjj>2.4$\TeV.
The $\chi^2/\mathrm{NDF}$ of the background-only fit, masking the signal region, with the ratio method is $42.04/32$ as shown
in Fig.~\ref{figDataAndFit}. The dijet mass spectrum is well modeled by both background prediction methods,
which also agree with one another.

\section{Limits on the resonance cross section, mass, and coupling}

We use the dijet mass spectrum from wide jets, the background parameterizations, and the dijet resonance shapes shown previously
to set limits on the production cross sections of new particles decaying to the parton pairs $\PQq\PQq$ (or $\PQq\PAQq$), $\Pq\Pg$, and $\Pg\Pg$.
A separate limit is determined for each final state ($\PQq\PQq$, $\Pq\Pg$, and $\Pg\Pg$) because of the dependence of the
dijet resonance shape on the types of the two final-state partons.

The dominant sources of systematic uncertainty are the jet energy scale and resolution,
the integrated luminosity, and the values of the parameters within the functional form
modeling the background shape in the dijet mass distribution. 
The uncertainty in the jet energy scale is within \jecUncert for all values of the dijet mass and is
determined from $\sqrt{s}=13$\TeV data using the methods described in Ref.~\cite{Khachatryan:2016kdb}.
This uncertainty is propagated to the limits by shifting the dijet mass shape for the signal by $\pm$\jecUncert.
The uncertainty in the jet energy resolution translates into an uncertainty of 10\% in the resolution of 
the dijet mass~\cite{Khachatryan:2016kdb}, and is propagated to the limits by observing the effect of 
increasing and decreasing by 10\% the
reconstructed width of the dijet mass shape for the signal.
The uncertainty in the integrated luminosity is 2.5\% in 2016~\cite{CMS-PAS-LUM-17-001} and 2018~\cite{CMS-PAS-LUM-18-002}, and \lumiUncert in 2017~\cite{CMS-PAS-LUM-17-004},
and is propagated to the  normalization of the signal.
Changes in the values of the parameters describing the background introduce a change in the signal yield,
which is accounted for as a systematic uncertainty, as discussed in the next paragraph.

The modified frequentist criterion~\cite{Junk1999,bib-cls} is
used to set upper limits on signal cross sections, following the prescription
described in Refs.~\cite{ATLAS:1379837,Cowan:2010js} using the asymptotic approximation of the test statistic.
We use a multi-bin counting experiment likelihood, which is
a product of Poisson distributions corresponding to different bins.
We evaluate the likelihood independently at each value of resonance pole mass from 1.8 to 8.7\TeV in 100-\GeVns steps.
The fit method is used to estimate the background for resonance masses from 1.8 to 2.9\TeV, and the ratio method
is used for resonance masses from 3.0 to 8.7\TeV . The minimum values of resonance mass
for the two methods, 1.8 and 3.0\TeV , are chosen to maintain reasonable acceptances for the minimum
$\mjj$ requirements, 1.5 and 2.4\TeV, respectively. The sources of systematic uncertainty are implemented as nuisance parameters in the likelihood model, with Gaussian constraints for the jet
energy scale and resolution, and log-normal constraints for the integrated luminosity. The background systematic uncertainty,
as we described previously, is automatically evaluated via profiling and decreases as the
resonance mass increases.

\subsection{Narrow resonances}

\begin{figure*}[hbtp!]
  \centering
    \includegraphics[width=0.48\textwidth]{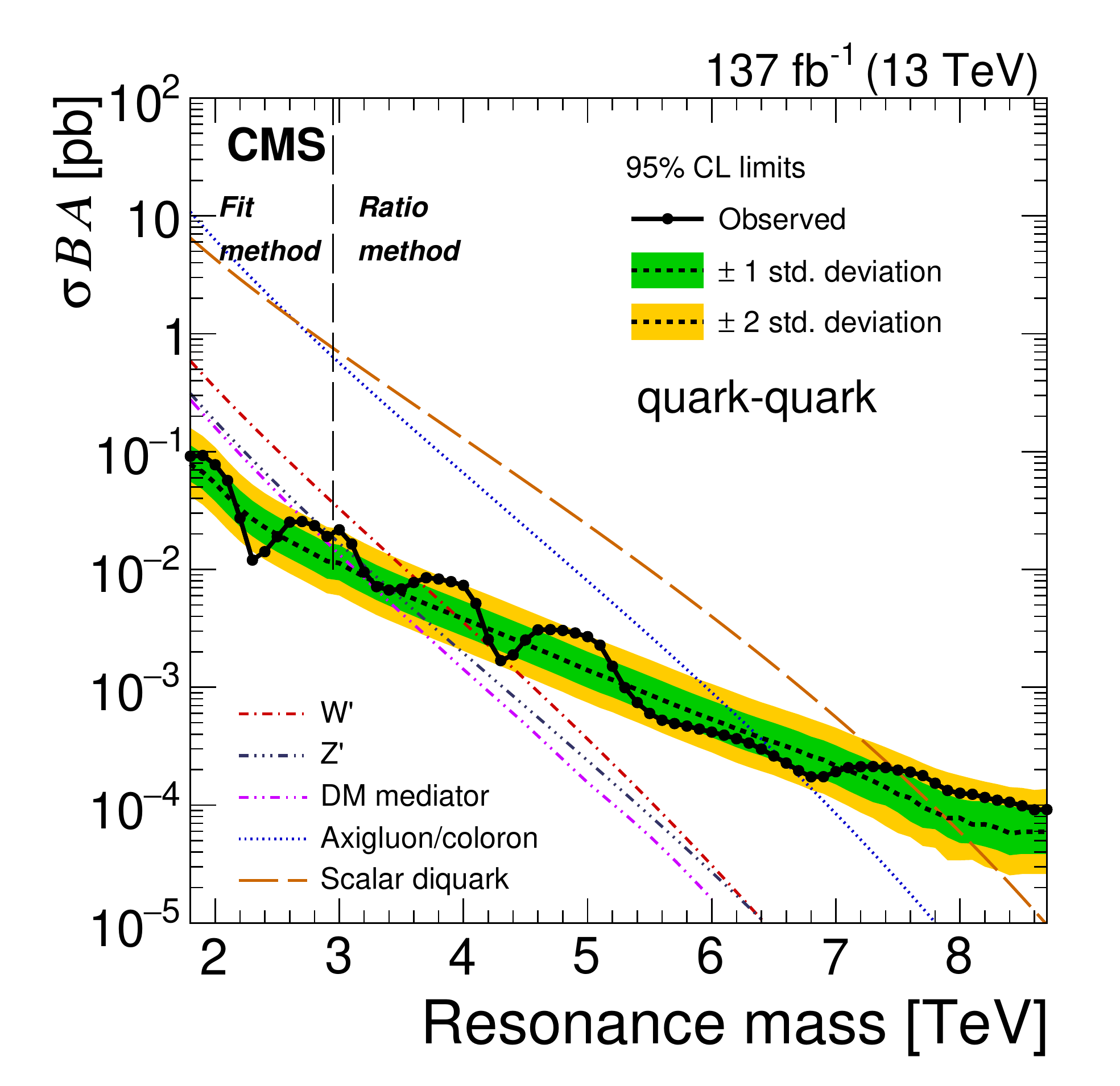}
    \includegraphics[width=0.48\textwidth]{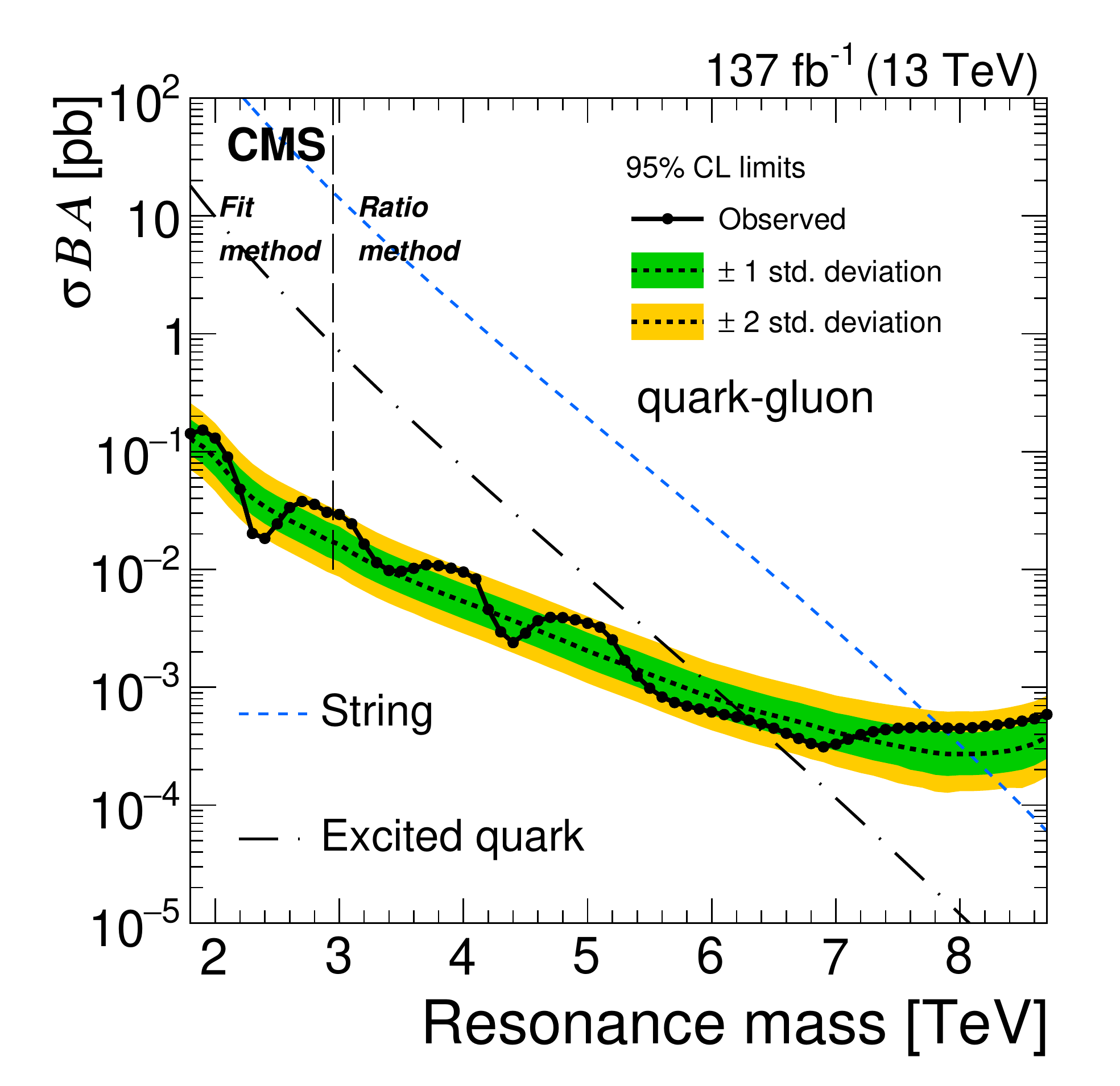}
    \includegraphics[width=0.48\textwidth]{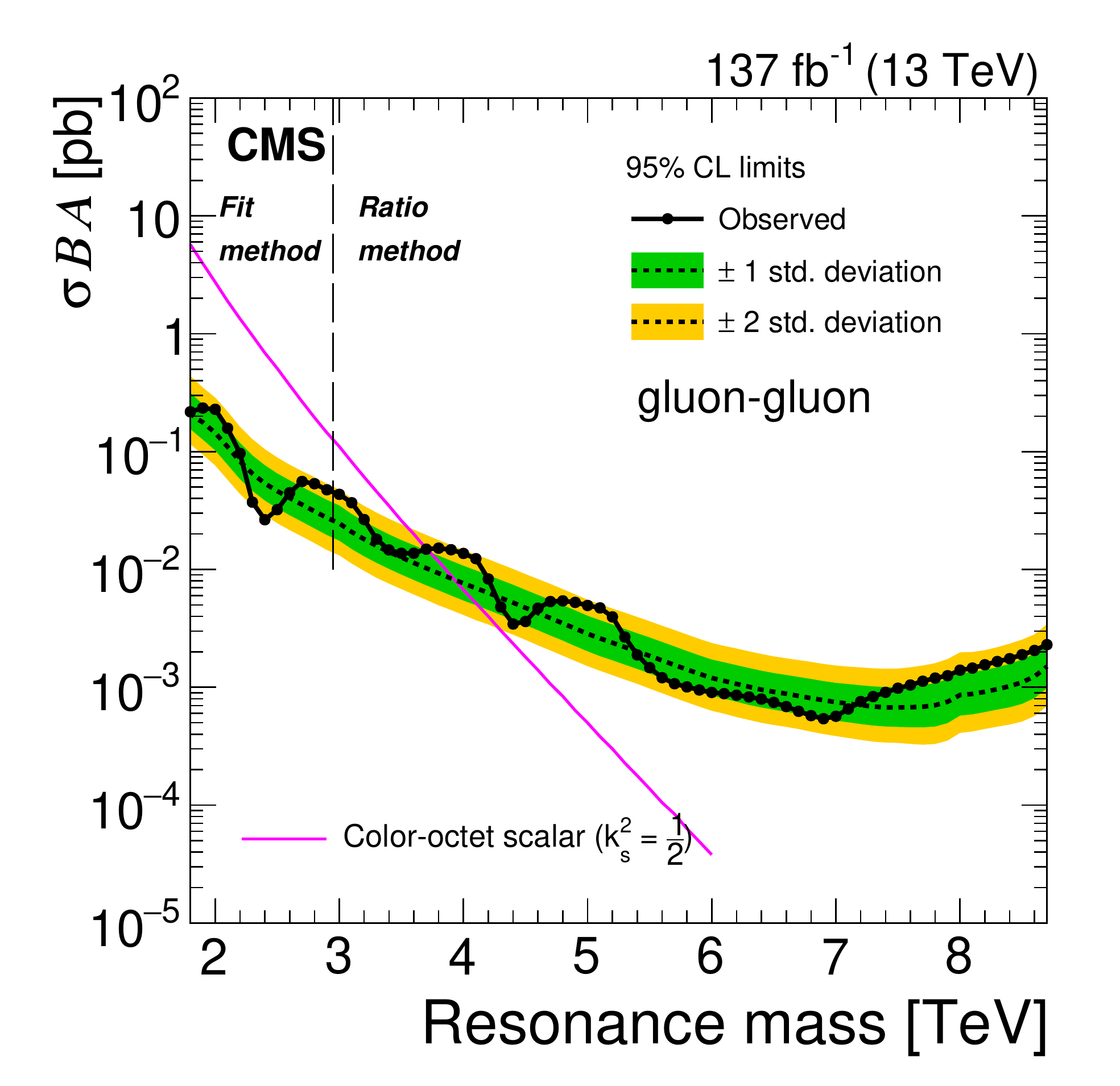}
    \includegraphics[width=0.48\textwidth]{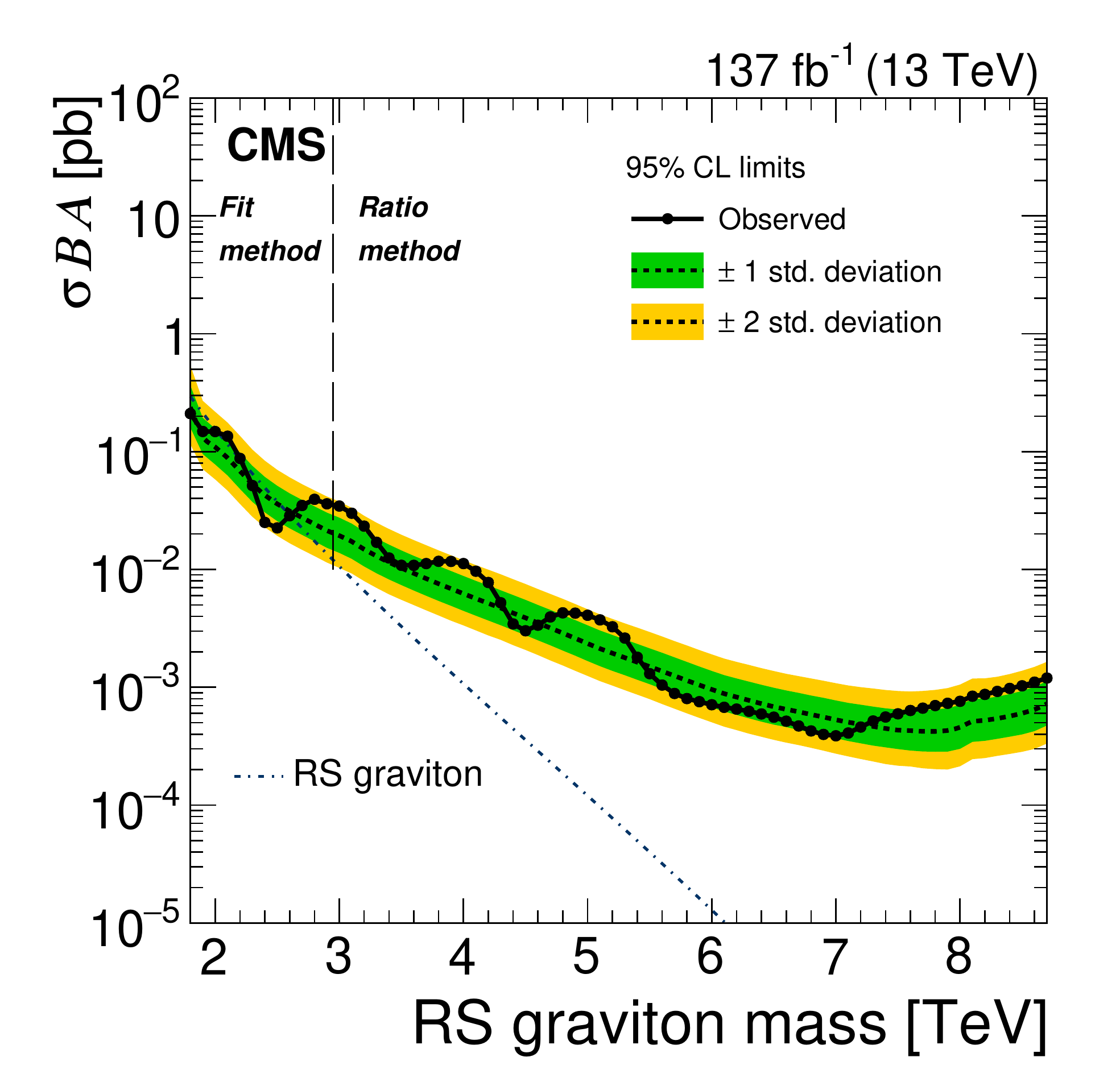}
    \caption{The observed 95\%~\CL upper limits on the product of the cross section, branching fraction, and acceptance for dijet resonances
    decaying to quark-quark (upper left), quark-gluon (upper right), gluon-gluon (lower left), and for RS gravitons (lower right).
    The corresponding expected limits (dashed lines) and their variations
    at the one and two standard deviation levels (shaded bands) are also shown.
   Limits are compared to predicted cross sections for string resonances~\cite{Anchordoqui:2008di,Cullen:2000ef},
    excited quarks~\cite{ref_qstar,Baur:1989kv}, axigluons~\cite{ref_axi}, colorons~\cite{ref_coloron},  scalar diquarks~\cite{ref_diquark},
    color-octet scalars~\cite{Han:2010rf}, new gauge bosons \PWpr and \PZpr with SM-like couplings~\cite{ref_gauge},
    DM mediators for $\mDM=1$\GeV~\cite{Boveia:2016mrp,Abdallah:2015ter}, and RS gravitons~\cite{ref_rsg}.
    The vertical dashed line indicates the boundary between the regions where
    the fit method and the ratio method are used to estimate the background.
    }
    \label{figLimitAll}
\end{figure*}

\begin{figure*}[hbtp!]
  \centering
    \includegraphics[width=0.98\textwidth]{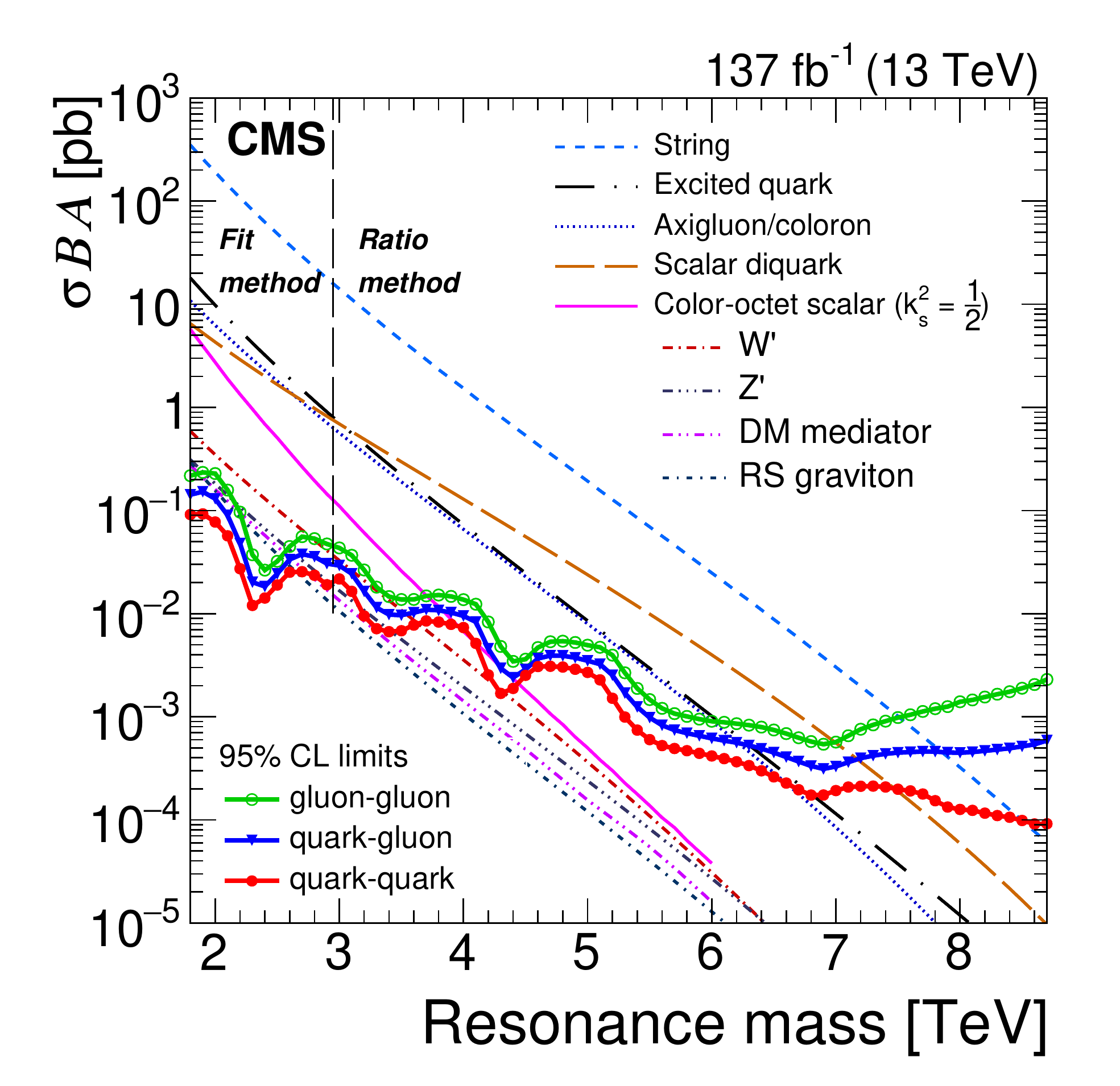}
    \caption{The observed 95\%~\CL upper limits on the product of the cross section, branching fraction, and acceptance for
    quark-quark, quark-gluon, and gluon-gluon type dijet resonances.
    Limits are compared to predicted cross sections for string resonances~\cite{Anchordoqui:2008di,Cullen:2000ef},
    excited quarks~\cite{ref_qstar,Baur:1989kv}, axigluons~\cite{ref_axi}, colorons~\cite{ref_coloron},  scalar diquarks~\cite{ref_diquark},
    color-octet scalars~\cite{Han:2010rf}, new gauge bosons \PWpr and \PZpr with SM-like couplings~\cite{ref_gauge},
    DM mediators for $\mDM=1$\GeV~\cite{Boveia:2016mrp,Abdallah:2015ter}, and RS gravitons~\cite{ref_rsg}.
    The vertical dashed line indicates the boundary between the regions where
    the fit method and the ratio method are used to estimate the background.
    }
    \label{figLimitSummary}
\end{figure*}

Figures~\ref{figLimitAll} and~\ref{figLimitSummary} show the model-independent observed
upper limits at 95\% confidence level on the product of
the cross section ($\sigma$), the branching fraction ($B$), and the acceptance ($A$) for narrow resonances, with the
kinematic requirements $\detajj<1.1$ for the dijet system and $\abs{\eta}<2.5$ for each jet.  The narrow resonance
shapes are the ones presented and discussed in detail in a previous publication~\cite{Sirunyan:2018xlo}.
The acceptance of the minimum dijet mass requirement in each search, which fully accounts for the overall experimental acceptance,
has been evaluated separately for $\PQq\PQq$, $\PQq\Pg$, and $\Pg\Pg$ resonances.
We include these acceptances in the determination of the limits. Figure \ref{figLimitAll} also shows the expected limits on $\sigma B A$ and their
bands of uncertainty.
Figure ~\ref{figLimitSummary}  shows the different limits for $\PQq\PQq$, $\PQq\Pg$, and $\Pg\Pg$ resonances, which
originate from differences in their line shapes. For the RS graviton, which decays to both $\PQq\PAQq$ and $\Pg\Pg$,
we obtain cross section upper limits from the average, weighted by branching fraction, of the limits on quark-quark
and gluon-gluon resonances.

\begin{figure}[!htb] \centering
\includegraphics[width=0.45\textwidth]{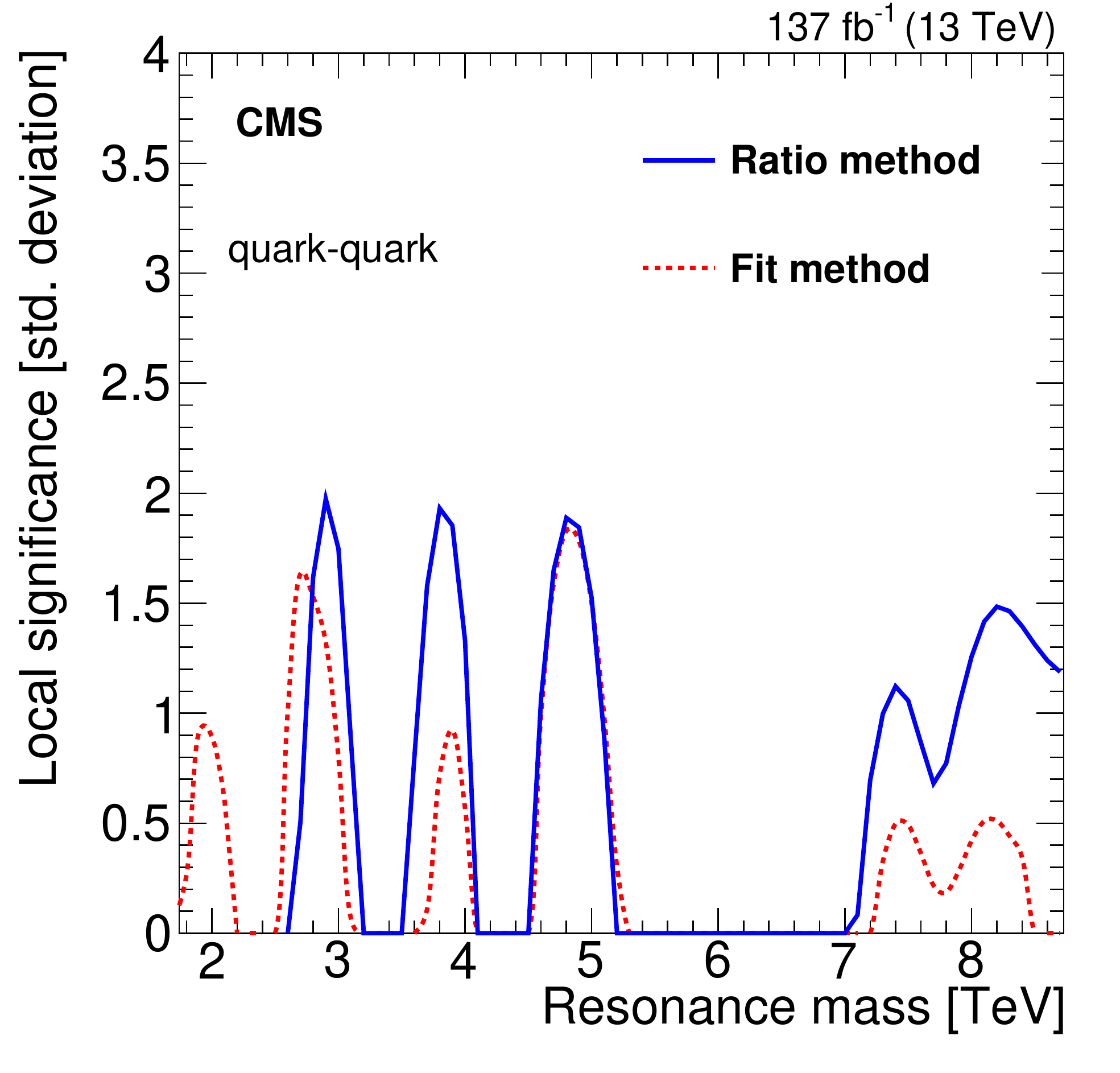}
\caption{Local significance for a $\Pq\Pq$ resonance with the ratio method (blue line) and the fit method (red dashed line).}
\label{fig:pfsignif}
\end{figure}

Using the statistical methodology discussed earlier, the local significance for $\PQq\PQq$, $\Pq\Pg$, and $\Pg\Pg$ resonance signals was measured from
1.8 to 8.7\TeV in steps of 100\GeV. The significance values obtained for $\PQq\PQq$ resonances are shown in
Fig.~\ref{fig:pfsignif} for both the ratio and the fit methods, and the significances for \Pq\Pg and \Pg\Pg resonances
are the same within 0.2 standard deviations. The ratio method usually gives a larger signal
significance than the fit method, because it provides a more accurate data-driven background
estimate.

All upper limits presented can be compared to the parton-level predictions of $\sigma\, B\, A$, without detector simulation,
to determine mass limits on new particles.
The model predictions shown in Fig.~\ref{figLimitAll}  are calculated in the narrow-width
approximation~\cite{Harris:2011bh} using the CTEQ6L1~\cite{refCTEQ} parton distribution function at LO.
An NLO correction factor of $K=1+ 8\pi\alpS/9\approx 1.3$ is applied to the LO predictions
for the \PWpr model and $K=1+(4\alpS/6\pi)(1+4\pi^2/3)\approx 1.2$ for the $\PZpr$  and the DM mediator models~\cite{Barger:1987nn}, where
\alpS is the
strong coupling constant evaluated at a scale equal to the resonance mass.
Similarly, for the axigluon and coloron models a correction factor is applied which varies between $K=1.1$ at a resonance mass of
$0.6$\TeV and $K=1.3$ at 8.1\TeV~\cite{Chivukula:2013xla}.
The branching fraction includes the direct decays of the resonance into the five light quarks and
gluons only, excluding top quarks from the decay, although top quarks are included in the calculation of
the resonance width.
The acceptance is evaluated at the parton level for the resonance decay to two partons. In the case of isotropic
decays, the acceptance is $A\approx 0.5$ and is independent of the resonance mass.
For a given model, new particles are excluded at 95\%~\CL in mass regions where the theoretical prediction
lies at or above the observed upper limit for the appropriate final state of Fig.~\ref{figLimitAll}.
Table~\ref{tab:MassLimit} shows the mass limits on all benchmark models which are extended
by 200 to 800\GeV relative to those reported in previous CMS dijet resonance searches~\cite{Sirunyan:2018xlo}.

\begin{table}[ht]
  \caption{ Observed and expected mass limits at 95\%~\CL from this analysis.
  The listed models are excluded between 1.8\TeV and the indicated mass limit by this analysis. The
  SM-like $\PZpr$ resonance is also excluded within the mass interval between 3.1 and 3.3\TeV.\\
}
\centering
\begin{tabular}{lcc}
Model    & Final state & Observed (expected) mass limit [\TeVns{}]                     \\ \hline
String   & $\PQq\Pg$                                                  & 7.9\ (8.1)   \\
Scalar diquark  & $\PQq\PQq$                                          & 7.5\ (7.9)   \\
Axigluon/coloron  & $\PQq\PAQq$                                       & 6.6\ (6.4)   \\
Excited quark  &  $\PQq\Pg$                                           & 6.3\ (6.2)   \\
Color-octet scalar ($k_s^2=1/2$) & $\Pg\Pg$                           & 3.7\ (3.9)   \\
\PWpr SM-like & $\PQq\PAQq$                                         & 3.6\ (3.9)   \\
\PZpr SM-like & $\PQq\PAQq$                                         & 2.9\ (3.4)   \\
RS graviton  ($k/\overline{M}_\text{Pl}=0.1$) & $\PQq\PAQq$, $\Pg\Pg$ & 2.6\ (2.6)   \\
DM mediator  ($\mDM=1$\GeV) & $\PQq\PAQq$                             & 2.8\ (3.2)   \\
\end{tabular}
\label{tab:MassLimit}
\end{table}

\subsection{Broad resonances}

\begin{figure}[!htb] \centering
\includegraphics[width=0.49\textwidth]{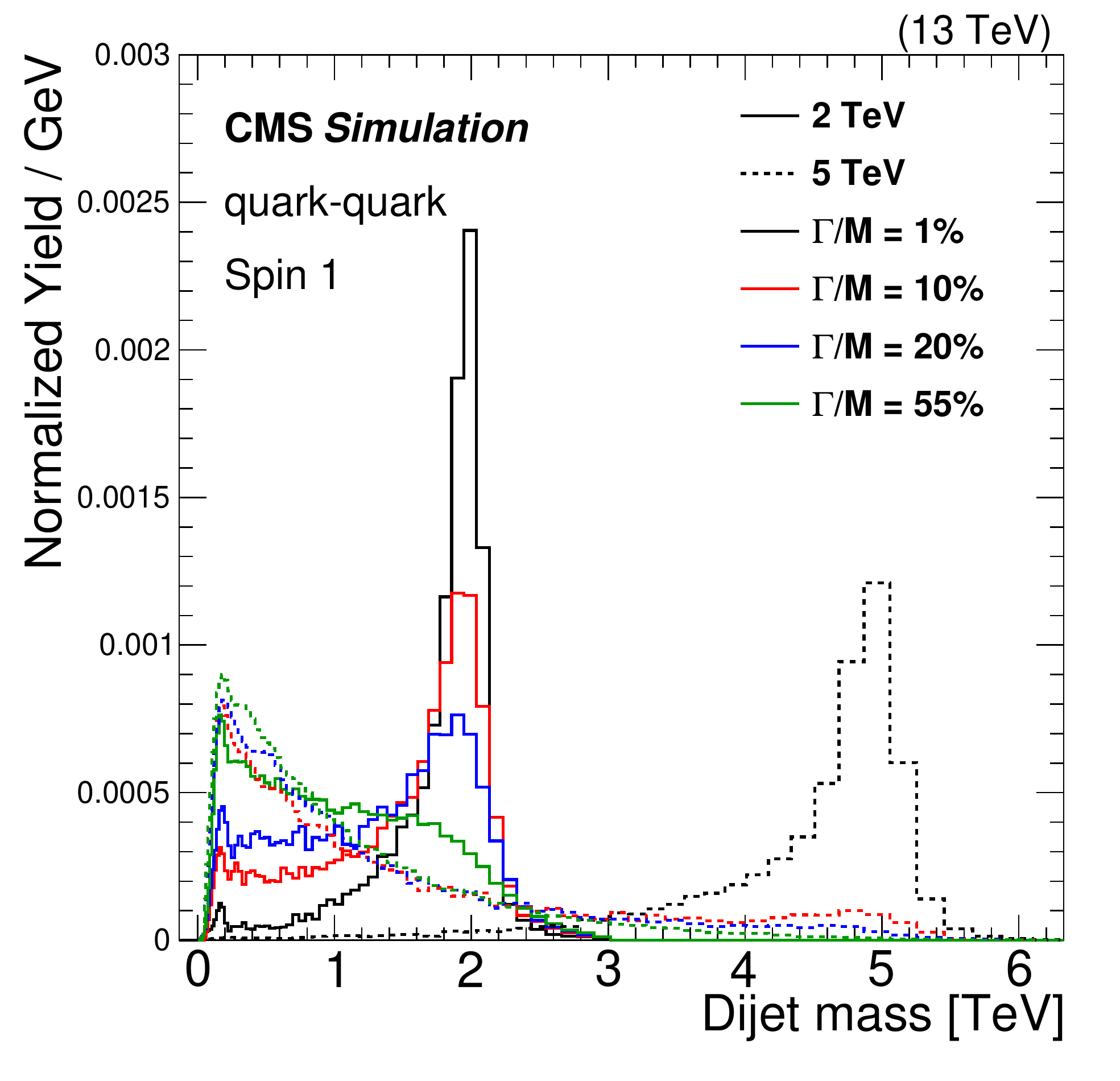}
\caption{The reconstructed dijet mass spectra for a vector particle decaying to pairs of quarks are shown for a resonance mass 
of 2\TeV (solid histogram) and 5\TeV (dashed histogram) for various 
values of intrinsic width, estimated from the \MADGRAPH{5} and \PYTHIA event generators followed by the simulation of the CMS detector response.}
\label{fig:wide_spin1_shape}
\end{figure}

We extend the search to cover broad resonances. We use spin-1  resonances decaying to quark-quark pairs with a width up to 55\% of the resonance mass, $M$, as  well as spin-2  resonances
that decay to quark or gluon pairs with a width up to 30\% of the resonance mass.
This allows us to be sensitive to more models and larger couplings. The spin-1 resonance results are also
used to produce  limits on the universal quark coupling of a leptophobic vector mediator of interactions between quarks and DM particles, and limits
for a leptophobic $\PZpr$ that couples to quarks but does not couple to DM particles~\cite{Chala:2015ama,Abercrombie:2015wmb,Abdallah:2015ter,Boveia:2016mrp}. In order to be sensitive to
the largest possible coupling values for these particles, the maximum value of examined
widths for spin-1 resonances is increased to  55\%  of the resonance mass.
The additional wider signals are produced in the same way as the narrower ones, using the \MGvATNLO
v. 2.3.2~\cite{Alwall:2014hca} generator at LO, and the \PYTHIA8.205~\cite{Sjostrand:2014zea} program,
followed  by a \GEANTfour-based \cite{refGEANT} simulation of the CMS detector.
For resonance widths up to 30\% of their mass, the dijet mass distributions are the ones presented and discussed in detail in Ref.~\cite{Sirunyan:2018xlo}.
The dijet mass distributions of both wide and narrow spin-1 resonances are shown in Fig.~\ref{fig:wide_spin1_shape}, and exhibit the same behavior
as the ones discussed in~\cite{Sirunyan:2018xlo}.

\begin{figure}[htb]
  \begin{center}
  \includegraphics[width=0.48\textwidth]{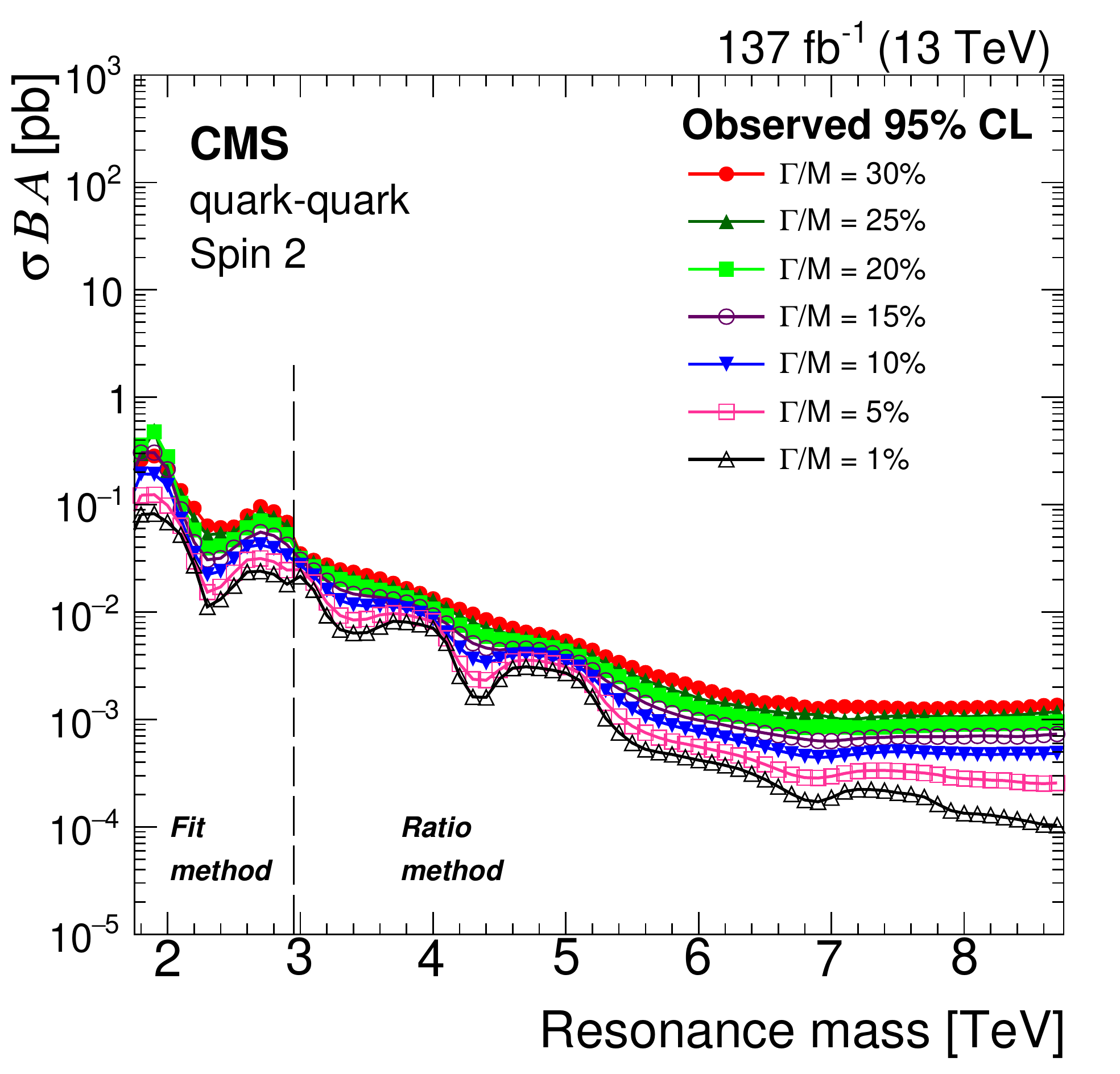}
  \includegraphics[width=0.48\textwidth]{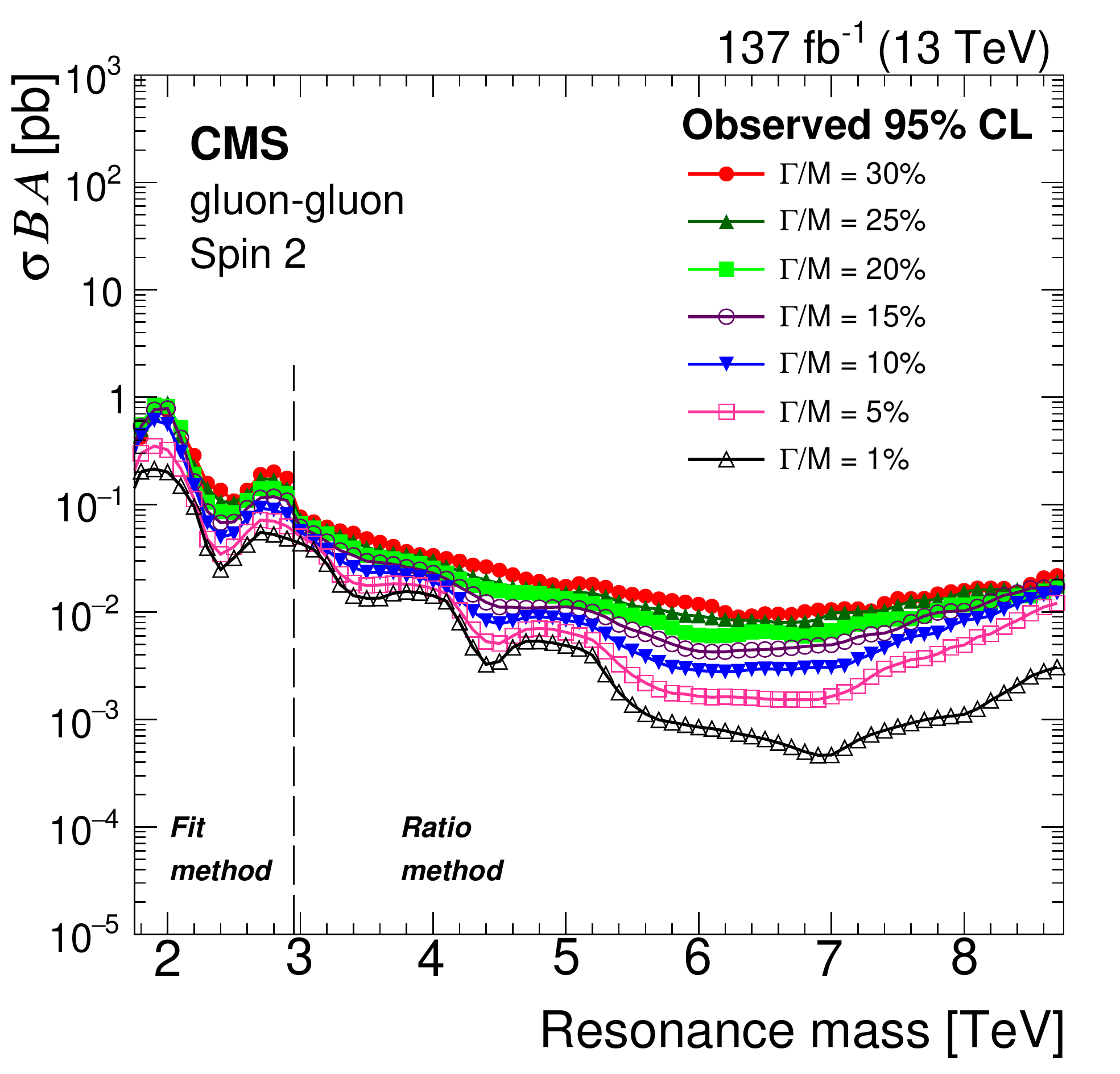}
  \includegraphics[width=0.48\textwidth]{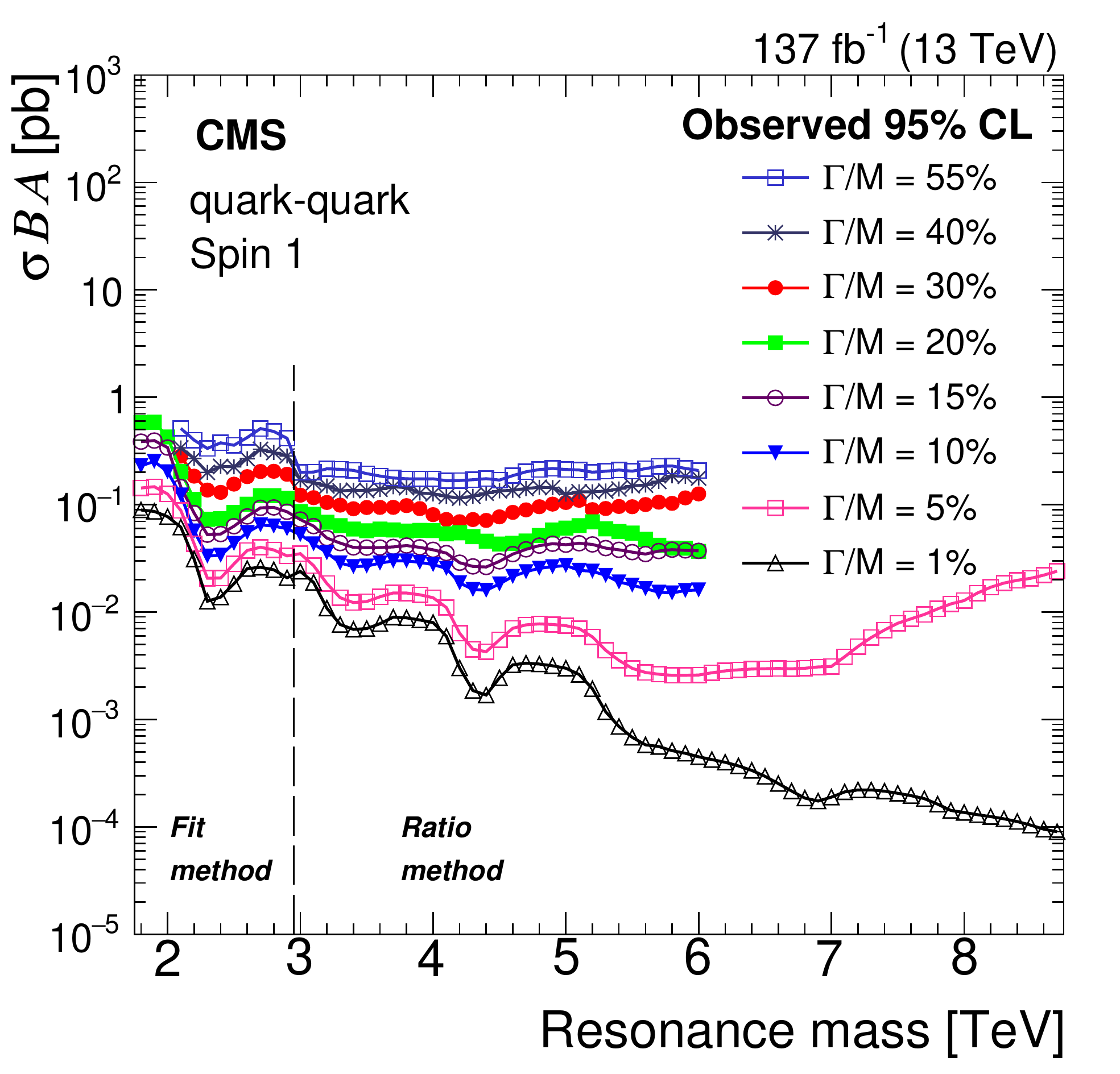}
  \caption{The observed 95\%~\CL upper limits on the product of the cross section, branching fraction, and acceptance
  for spin-2 resonances produced and decaying in the quark-quark (upper left) and gluon-gluon (upper right) channels, as well as for spin-1 resonances
  decaying in the quark-quark channel (lower), shown for various values of intrinsic width as a function of resonance mass.
   The vertical dashed line indicates the boundary between the regions where
   the fit method and the ratio method are used to estimate the background.
  }
  \label{fig:wide_qq}
  \end{center}
\end{figure}

The cross section limits in this case are presented as a function of resonance mass and width.
In Fig.~\ref{fig:wide_qq}  we show the observed 95\%~\CL upper limits
for various resonance widths, for spin-2 resonances modeled by an RS graviton signal in the quark-quark and
gluon-gluon channels, and for spin-1 resonances in the quark-quark channel.
The limits weaken as the resonance intrinsic width increases, following the characteristics of the resonance
shapes. The spin-1 resonances
are significantly broader than the spin-2 resonances. For this reason, their limits are weaker than those of the spin-2 resonances.
In Fig.~\ref{fig:wide_qq} the cross section limits at very high mass for spin-1 resonances with $\Gamma/M=5$\% increase as the resonance mass increases, while 
they decrease for  $\Gamma/M=1$\%. This is because, for resonances with widths larger than 1\%, the tail to low dijet mass increases significantly as the 
resonance 
mass increases, as shown in Fig.~\ref{fig:wide_spin1_shape}.

\begin{figure}[hbt]
   \centering
   \includegraphics[width=0.45\textwidth]{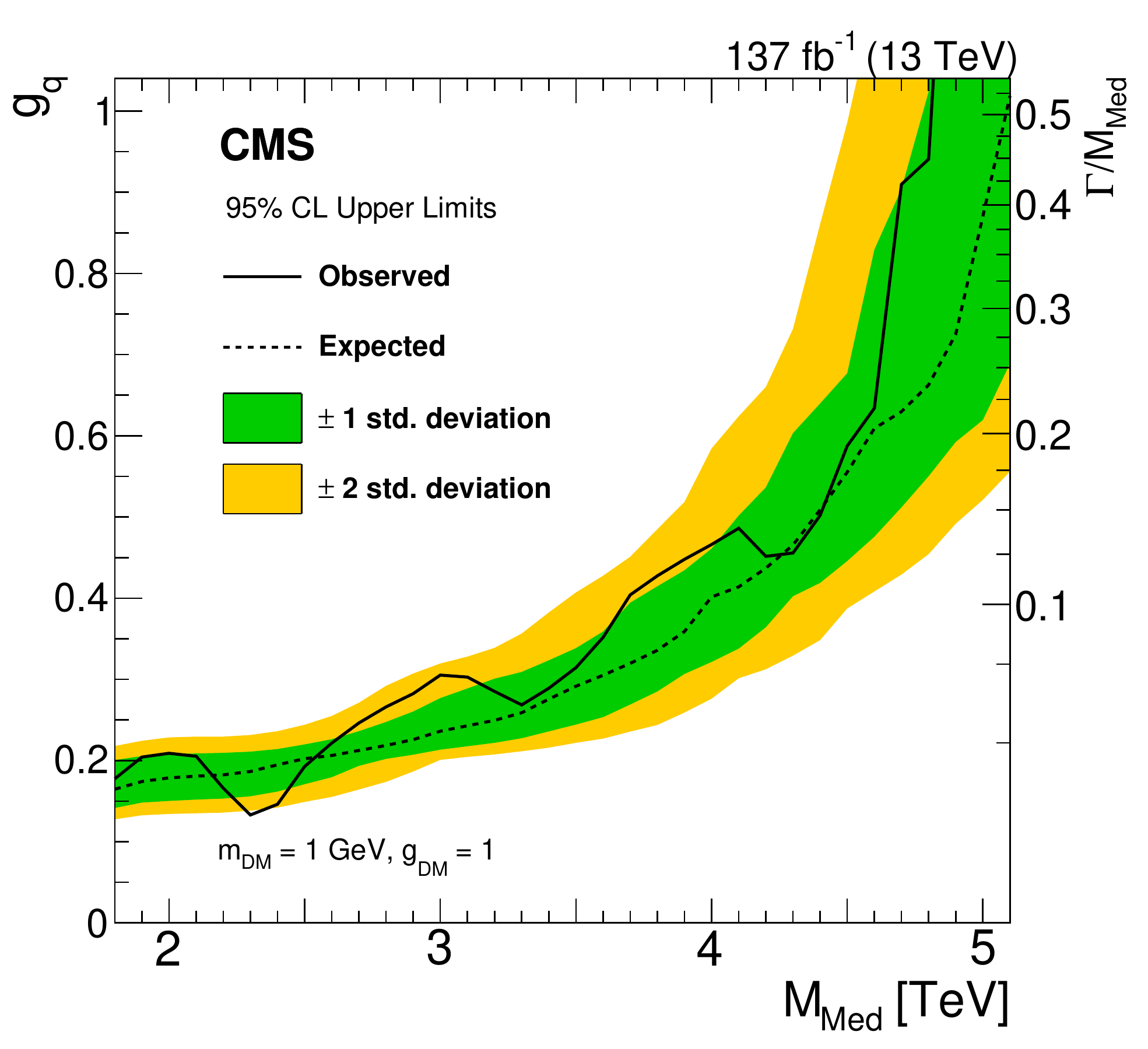}
   \includegraphics[width=0.45\textwidth]{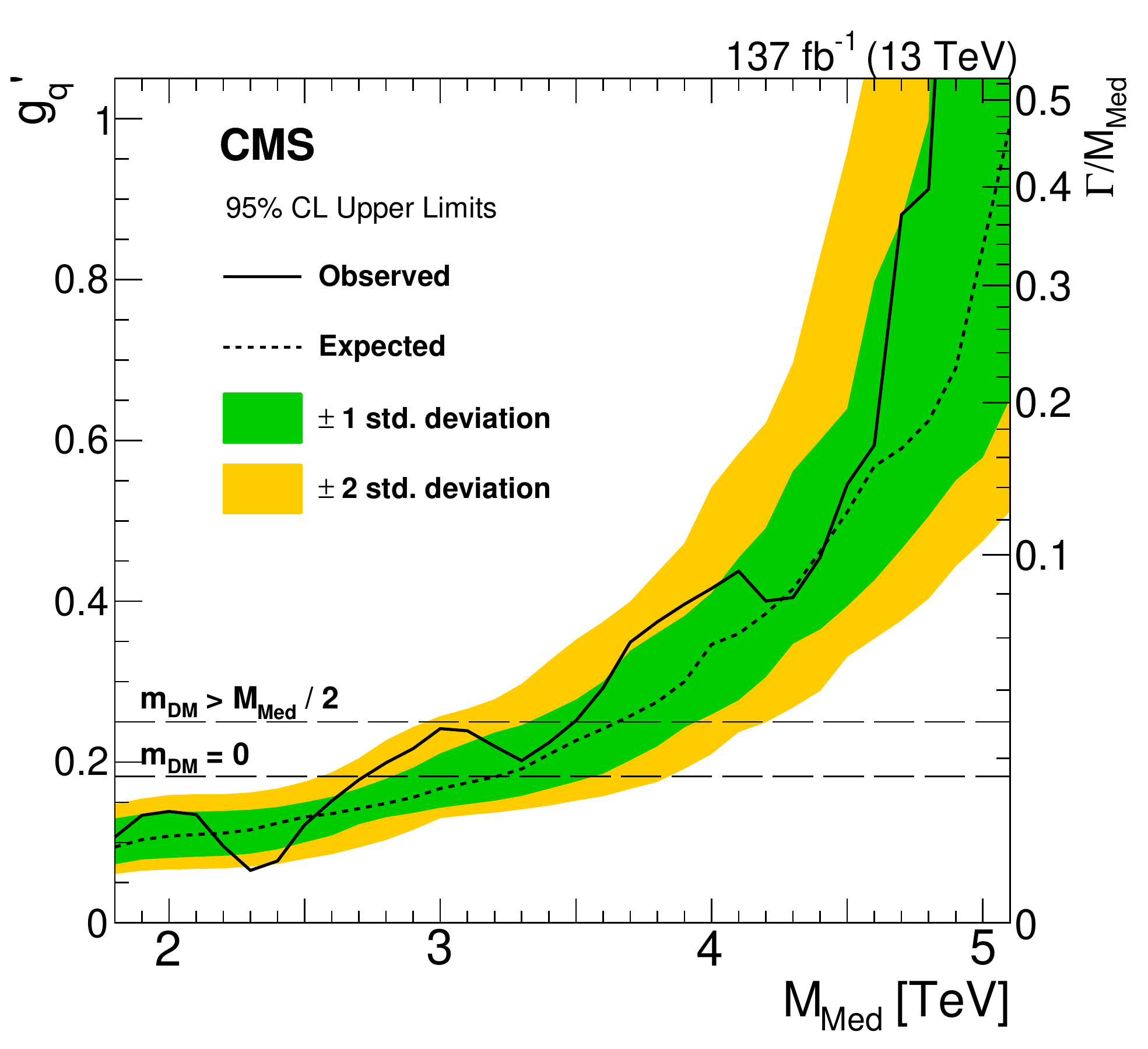}
   \caption{The 95\%~\CL upper limits on the universal quark coupling $\gq$ as a function of resonance mass for a vector mediator of
    interactions
    between quarks and DM particles (left), and between quarks only (right). The dashed horizontal lines on the right plot show the
    coupling strength
    for which the cross section for dijet production in this leptophobic $\PZpr$  model is the same as for a DM mediator for $\gq=0.25$.
    The right vertical axis shows the natural width of the mediator divided by its mass.
    The expected limits (dashed lines) and their variation at the one and two standard deviation levels (shaded bands) are also shown.}
    \label{figCouplingWide}
\end{figure}

The limits are presented up to a maximum resonance mass of 8.7\TeV for most models. We do not present limits for the case of spin-1 resonances in the
quark-quark channel with masses larger than 6\TeV and $\Gamma/M > 0.1$. These resonances are not part of the search because they have an
exceedingly broad and high tail at low dijet mass, as described in Ref.~\cite{Sirunyan:2018xlo}, which dominates the limit and
produces unstable search results.  The spin-1 cross section limits in Fig.~\ref{fig:wide_qq} have been used to derive constraints on the coupling to quarks of mediators of new interactions. We
consider two models of a leptophobic  mediator which  couples to all generations of quarks with the same universal strength. The quark coupling is denoted $\gq^{\prime}$ in the first model, in which
the mediator does not couple to DM  particles, and denoted $\gq$ in the second model, in which the mediator couples to DM particles.
For each mediator mass value, the predictions for the cross section of mediator production as
a function of the quark coupling are converted to predictions as a function of width. They are then compared to the spin-1 cross section limits
from Fig.~\ref{fig:wide_qq} to find the excluded values of quark coupling, as a function of mass for a spin-1 resonance.
Figure~\ref{figCouplingWide} (right) shows upper limits on the coupling $\gq^{\prime}$ as a function of mass
for our first model, also known as a leptophobic  $\PZpr$ resonance~\cite{Dobrescu:2013coa} that couples only to quarks. In this model the resonance has a width
\begin{equation}
\Gamma_{\text{Med}} = \frac{3(\gq^{\prime})^2 \mMed}{2\pi},
\label{eqWidthZp}
\end{equation}
where $\mMed$ is the resonance mass and $\gq^{\prime}$ is the universal quark coupling, related to the
coupling, $g_{B}$, of Ref.~\cite{Dobrescu:2013coa} by $\gq'=g_{B}/6$.
Figure~\ref{figCouplingWide} (left) shows upper limits on the coupling  $\gq$  as a function of mass
for our second model, also known as a DM Mediator model, which has a leptophobic spin-1 mediator that couples
both to quarks and DM particles~\cite{Boveia:2016mrp}, and for Dirac DM with a mass $\mDM=1$\GeV and a coupling $\gDM=1.0$.
The cross section of mediator production for $\mDM=1$\GeV and $\gDM=1$ is calculated with \MGvATNLO
~\cite{Alwall:2014hca} for mediator masses within the range $1.6<\mMed<5.1$\TeV in $0.1$\TeV steps and for quark couplings
within the range $0.1<\gq<1.0$ in $0.1$ steps. For these choices, the relationship between the total mediator width, for decays to both quark and DM particles,
and  $\gq$ given in Refs.~\cite{Boveia:2016mrp,Abdallah:2015ter} simplifies to
\begin{equation}
\Gamma_{\text{Med}} \approx \frac{(18\gq^2 + 1)\mMed}{12\pi}.
\label{eqWidth}
\end{equation}

The increased sensitivity of the ratio method to wide resonances significantly improves and extends previous limits on DM mediators at large values of  $\Gamma/M$.
For example, for $\Gamma/M=0.45$, this search excludes DM mediators with mass less
than 4.8\TeV, while the observed limit from the earlier searches was 4.0\TeV~\cite{Sirunyan:2018xlo}.

\section{Summary}

A search for resonances decaying into a pair of jets has been performed using proton-proton collision data at $\sqrt{s}=13$\TeV corresponding to an integrated
luminosity of \RunLumi. The dijet mass spectra are observed to be smoothly falling distributions of events with typically  two-jet topology,
although one unusual event with a four-jet topology was found at high mass. The background is predicted using two methods.
The fit method uses an empirical functional form to fit the background in the signal region, defined by requiring the pseudorapidity separation of two jets in dijet $\abs{\Delta\eta}<$1.1, while
the ratio method
uses two control regions at higher values of
$\abs{\Delta\eta}$ to predict the background in the signal region. The ratio method is a new background prediction method, which is independent of and
complementary to the fit method. No evidence for resonant particle production is observed. Generic upper limits are presented on
the product of the cross section, the branching fraction, and  the acceptance for narrow and broad quark-quark,
quark-gluon, and gluon-gluon resonances. The limits are applied to various models of new resonances and yield the following 95\% confidence level lower limits on the resonance masses:
7.9\TeV for string resonances,
7.5\TeV for scalar diquarks, 6.6\TeV for axigluons and colorons, 6.3\TeV for excited quarks,  3.7\TeV for color-octet
scalars, 3.6\TeV for \PWpr bosons with SM-like couplings, 2.9\TeV and between 3.1
and 3.3\TeV for $\PZpr$ bosons with SM-like couplings,  2.6\TeV for Randall--Sundrum gravitons, and 2.8\TeV for dark matter (DM) mediators.
With this search, limits on narrow resonances are improved by 200 to 800\GeV relative to those reported in previous CMS dijet resonance
searches. Limits are also presented for spin-2 resonances with intrinsic widths as large as 30\% of the resonance mass, and spin-1 resonances with
intrinsic widths as large as  55\% of the resonance
mass. These limits are used to improve and extend  the exclusions of a DM mediator to larger values of the resonance mass and coupling to quarks.
In the search for broad resonances, the ratio method provides significantly enhanced sensitivity compared to the fit method,
resulting in the exclusion at 95\% confidence level of a DM mediator with mass less than
4.8\TeV for a width equal to 45\% of the mass, which corresponds to a coupling to quarks $\gq=0.9$.

\begin{acknowledgments}
We congratulate our colleagues in the CERN accelerator departments for the excellent performance of the LHC and thank the technical and administrative staffs at CERN and at other CMS institutes for their contributions to the success of the CMS effort. In addition, we gratefully acknowledge the computing centers and personnel of the Worldwide LHC Computing Grid for delivering so effectively the computing infrastructure essential to our analyses. Finally, we acknowledge the enduring support for the construction and operation of the LHC and the CMS detector provided by the following funding agencies: BMBWF and FWF (Austria); FNRS and FWO (Belgium); CNPq, CAPES, FAPERJ, FAPERGS, and FAPESP (Brazil); MES (Bulgaria); CERN; CAS, MoST, and NSFC (China); COLCIENCIAS (Colombia); MSES and CSF (Croatia); RPF (Cyprus); SENESCYT (Ecuador); MoER, ERC IUT, PUT and ERDF (Estonia); Academy of Finland, MEC, and HIP (Finland); CEA and CNRS/IN2P3 (France); BMBF, DFG, and HGF (Germany); GSRT (Greece); NKFIA (Hungary); DAE and DST (India); IPM (Iran); SFI (Ireland); INFN (Italy); MSIP and NRF (Republic of Korea); MES (Latvia); LAS (Lithuania); MOE and UM (Malaysia); BUAP, CINVESTAV, CONACYT, LNS, SEP, and UASLP-FAI (Mexico); MOS (Montenegro); MBIE (New Zealand); PAEC (Pakistan); MSHE and NSC (Poland); FCT (Portugal); JINR (Dubna); MON, RosAtom, RAS, RFBR, and NRC KI (Russia); MESTD (Serbia); SEIDI, CPAN, PCTI, and FEDER (Spain); MOSTR (Sri Lanka); Swiss Funding Agencies (Switzerland); MST (Taipei); ThEPCenter, IPST, STAR, and NSTDA (Thailand); TUBITAK and TAEK (Turkey); NASU (Ukraine); STFC (United Kingdom); DOE and NSF (USA).

\hyphenation{Rachada-pisek} Individuals have received support from the Marie-Curie program and the European Research Council and Horizon 2020 Grant, contract Nos.\ 675440, 752730, and 765710 (European Union); the Leventis Foundation; the A.P.\ Sloan Foundation; the Alexander von Humboldt Foundation; the Belgian Federal Science Policy Office; the Fonds pour la Formation \`a la Recherche dans l'Industrie et dans l'Agriculture (FRIA-Belgium); the Agentschap voor Innovatie door Wetenschap en Technologie (IWT-Belgium); the F.R.S.-FNRS and FWO (Belgium) under the ``Excellence of Science -- EOS" -- be.h project n.\ 30820817; the Beijing Municipal Science \& Technology Commission, No. Z181100004218003; the Ministry of Education, Youth and Sports (MEYS) of the Czech Republic; the Lend\"ulet (``Momentum") Program and the J\'anos Bolyai Research Scholarship of the Hungarian Academy of Sciences, the New National Excellence Program \'UNKP, the NKFIA research grants 123842, 123959, 124845, 124850, 125105, 128713, 128786, and 129058 (Hungary); the Council of Science and Industrial Research, India; the HOMING PLUS program of the Foundation for Polish Science, cofinanced from European Union, Regional Development Fund, the Mobility Plus program of the Ministry of Science and Higher Education, the National Science Center (Poland), contracts Harmonia 2014/14/M/ST2/00428, Opus 2014/13/B/ST2/02543, 2014/15/B/ST2/03998, and 2015/19/B/ST2/02861, Sonata-bis 2012/07/E/ST2/01406; the National Priorities Research Program by Qatar National Research Fund; the Ministry of Science and Education, grant no. 3.2989.2017 (Russia); the Programa Estatal de Fomento de la Investigaci{\'o}n Cient{\'i}fica y T{\'e}cnica de Excelencia Mar\'{\i}a de Maeztu, grant MDM-2015-0509 and the Programa Severo Ochoa del Principado de Asturias; the Thalis and Aristeia programs cofinanced by EU-ESF and the Greek NSRF; the Rachadapisek Sompot Fund for Postdoctoral Fellowship, Chulalongkorn University and the Chulalongkorn Academic into Its 2nd Century Project Advancement Project (Thailand); the Nvidia Corporation; the Welch Foundation, contract C-1845; and the Weston Havens Foundation (USA).
\end{acknowledgments}

\bibliography{auto_generated}

\providecommand{\href}[2]{#2}\begingroup\raggedright\begin{thebibliography}{10}%
\makeatletter
\providecommand{\hrefCMSnoop }[0]{\@secondoftwo}%
\makeatother
\providecommand{\doi}{\texttt{doi:}\begingroup \urlstyle{tt}\Url}

\bibitem{Anchordoqui:2008di}
L.~A. Anchordoqui\hrefCMSnoop {}{ {et~al.}, ``Dijet signals for low mass
  strings at the {LHC}'',} \textit{ Phys. Rev. Lett.} \textbf{ 101} (2008)
  241803,
  \href{http://dx.doi.org/10.1103/PhysRevLett.101.241803}{\doi{10.1103/PhysRevLett.101.241803}},
\href{http://www.arXiv.org/abs/0808.0497}{\texttt{arXiv:0808.0497}}.
%%CITATION = ARXIV:0808.0497;%%.

\bibitem{Cullen:2000ef}
\hrefCMSnoop {}{S.~Cullen, M.~Perelstein, and M.~E. Peskin, ``{TeV strings and
  collider probes of large extra dimensions}'',} \textit{ Phys. Rev. D}
  \textbf{ 62} (2000) 055012,
  \href{http://dx.doi.org/10.1103/PhysRevD.62.055012}{\doi{10.1103/PhysRevD.62.055012}},
\href{http://www.arXiv.org/abs/hep-ph/0001166}{\texttt{arXiv:hep-ph/0001166}}.
%%CITATION = HEP-PH/0001166;%%.

\bibitem{ref_diquark}
\hrefCMSnoop {}{J.~L. Hewett and T.~G. Rizzo, ``Low-energy phenomenology of
  superstring-inspired {E(6)} models'',} \textit{ Phys. Rept.} \textbf{ 183}
  (1989) 193,
\href{http://dx.doi.org/10.1016/0370-1573(89)90071-9}{\doi{10.1016/0370-1573(89)90071-9}}.
%%CITATION = PRPLC,183,193;%%.

\bibitem{ref_qstar}
\hrefCMSnoop {}{U.~Baur, I.~Hinchliffe, and D.~Zeppenfeld, ``Excited quark
  production at hadron colliders'',} \textit{ Int. J. Mod. Phys. A} \textbf{
  02} (1987) 1285,
\href{http://dx.doi.org/10.1142/S0217751X87000661}{\doi{10.1142/S0217751X87000661}}.
%%CITATION = IMPAE,A2,1285;%%.

\bibitem{Baur:1989kv}
\hrefCMSnoop {}{U.~Baur, M.~Spira, and P.~M. Zerwas, ``Excited quark and lepton
  production at hadron colliders'',} \textit{ Phys. Rev. D} \textbf{ 42} (1990)
  815,
\href{http://dx.doi.org/10.1103/PhysRevD.42.815}{\doi{10.1103/PhysRevD.42.815}}.
%%CITATION = PHRVA,D42,815;%%.

\bibitem{ref_axi}
\hrefCMSnoop {}{P.~H. Frampton and S.~L. Glashow, ``Chiral color: An
  alternative to the standard model'',} \textit{ Phys. Lett. B} \textbf{ 190}
  (1987) 157,
\href{http://dx.doi.org/10.1016/0370-2693(87)90859-8}{\doi{10.1016/0370-2693(87)90859-8}}.
%%CITATION = PHLTA,B190,157;%%.

\bibitem{Chivukula:2013xla}
\hrefCMSnoop {}{R.~S. Chivukula, E.~H. Simmons, A.~Farzinnia, and J.~Ren,
  ``Hadron collider production of massive color-octet vector bosons at
  next-to-leading order'',} \textit{ Phys. Rev. D} \textbf{ 87} (2013) 094011,
  \href{http://dx.doi.org/10.1103/PhysRevD.87.094011}{\doi{10.1103/PhysRevD.87.094011}},
\href{http://www.arXiv.org/abs/1303.1120}{\texttt{arXiv:1303.1120}}.
%%CITATION = ARXIV:1303.1120;%%.

\bibitem{ref_coloron}
\hrefCMSnoop {}{E.~H. Simmons, ``Coloron phenomenology'',} \textit{ Phys. Rev.
  D} \textbf{ 55} (1997) 1678,
  \href{http://dx.doi.org/10.1103/PhysRevD.55.1678}{\doi{10.1103/PhysRevD.55.1678}},
\href{http://www.arXiv.org/abs/hep-ph/9608269}{\texttt{arXiv:hep-ph/9608269}}.
%%CITATION = HEP-PH/9608269;%%.

\bibitem{Han:2010rf}
\hrefCMSnoop {}{T.~Han, I.~Lewis, and Z.~Liu, ``Colored resonant signals at the
  {LHC}: largest rate and simplest topology'',} \textit{ JHEP} \textbf{ 12}
  (2010) 085,
  \href{http://dx.doi.org/10.1007/JHEP12(2010)085}{\doi{10.1007/JHEP12(2010)085}},
\href{http://www.arXiv.org/abs/1010.4309}{\texttt{arXiv:1010.4309}}.
%%CITATION = ARXIV:1010.4309;%%.

\bibitem{ref_gauge}
\hrefCMSnoop {}{E.~Eichten, I.~Hinchliffe, K.~D. Lane, and C.~Quigg,
  ``Supercollider physics'',} \textit{ Rev. Mod. Phys.} \textbf{ 56} (1984)
  579,
\href{http://dx.doi.org/10.1103/RevModPhys.56.579}{\doi{10.1103/RevModPhys.56.579}}.
%%CITATION = RMPHA,56,579;%%.

\bibitem{ref_rsg}
\hrefCMSnoop {}{L.~Randall and R.~Sundrum, ``An alternative to
  compactification'',} \textit{ Phys. Rev. Lett.} \textbf{ 83} (1999) 4690,
  \href{http://dx.doi.org/10.1103/PhysRevLett.83.4690}{\doi{10.1103/PhysRevLett.83.4690}},
\href{http://www.arXiv.org/abs/hep-th/9906064}{\texttt{arXiv:hep-th/9906064}}.
%%CITATION = HEP-TH/9906064;%%.

\bibitem{Chala:2015ama}
M.~Chala\hrefCMSnoop {}{ {et~al.}, ``Constraining dark sectors with monojets
  and dijets'',} \textit{ JHEP} \textbf{ 07} (2015) 089,
  \href{http://dx.doi.org/10.1007/JHEP07(2015)089}{\doi{10.1007/JHEP07(2015)089}},
\href{http://www.arXiv.org/abs/1503.05916}{\texttt{arXiv:1503.05916}}.
%%CITATION = ARXIV:1503.05916;%%.

\bibitem{Abercrombie:2015wmb}
\hrefCMSnoop {}{D.~Abercrombie {et~al.}, ``Dark matter benchmark models for
  early {LHC} run-2 searches: Report of the {ATLAS/CMS} dark matter forum'',}
  \textit{ Phys. Dark Univ.} \textbf{ 26} (2019) 100371,
  \href{http://dx.doi.org/10.1016/j.dark.2019.100371}{\doi{10.1016/j.dark.2019.100371}},
\href{http://www.arXiv.org/abs/1507.00966}{\texttt{arXiv:1507.00966}}.
%%CITATION = ARXIV:1507.00966;%%.

\bibitem{Abdallah:2015ter}
\hrefCMSnoop {}{J.~Abdallah {et~al.}, ``Simplified models for dark matter
  searches at the {LHC}'',} \textit{ Phys. Dark Univ.} \textbf{ 9-10} (2015) 8,
  \href{http://dx.doi.org/10.1016/j.dark.2015.08.001}{\doi{10.1016/j.dark.2015.08.001}},
\href{http://www.arXiv.org/abs/1506.03116}{\texttt{arXiv:1506.03116}}.
%%CITATION = ARXIV:1506.03116;%%.

\bibitem{Boveia:2016mrp}
\hrefCMSnoop {}{G.~Busoni {et~al.}, ``{Recommendations on presenting LHC
  searches for missing transverse energy signals using simplified $s$-channel
  models of dark matter}'',} (2016).
\href{http://www.arXiv.org/abs/1603.04156}{\texttt{arXiv:1603.04156}}.
%%CITATION = ARXIV:1603.04156;%%.

\bibitem{Aad:2019hjw}
\hrefCMSnoop {}{{ATLAS Collaboration}, ``{Search for new resonances in mass
  distributions of jet pairs using 139 fb$^{-1}$ of $pp$ collisions at
  $\sqrt{s}=13$ TeV with the ATLAS detector}'',} (2019).
  \href{http://www.arXiv.org/abs/1910.08447}{\texttt{arXiv:1910.08447}}.
Submitted to \textit{JHEP}.
%%CITATION = ARXIV:1910.08447;%%.

\bibitem{Sirunyan:2018xlo}
\hrefCMSnoop {}{{CMS Collaboration}, ``{Search for narrow and broad dijet
  resonances in proton-proton collisions at $\sqrt{s}=$ 13 TeV and constraints
  on dark matter mediators and other new particles}'',} \textit{ JHEP} \textbf{
  08} (2018) 130,
  \href{http://dx.doi.org/10.1007/JHEP08(2018)130}{\doi{10.1007/JHEP08(2018)130}},
\href{http://www.arXiv.org/abs/1806.00843}{\texttt{arXiv:1806.00843}}.
%%CITATION = ARXIV:1806.00843;%%.

\bibitem{Aaboud:2017yvp}
\hrefCMSnoop {}{{ATLAS Collaboration}, ``{Search for new phenomena in dijet
  events using 37 fb$^{-1}$ of $pp$ collision data collected at $\sqrt{s}=$13
  TeV with the ATLAS detector}'',} \textit{ Phys. Rev. D} \textbf{ 96} (2017)
  052004,
  \href{http://dx.doi.org/10.1103/PhysRevD.96.052004}{\doi{10.1103/PhysRevD.96.052004}},
\href{http://www.arXiv.org/abs/1703.09127}{\texttt{arXiv:1703.09127}}.
%%CITATION = ARXIV:1703.09127;%%.

\bibitem{Sirunyan:2016iap}
\hrefCMSnoop {}{{CMS Collaboration}, ``{Search for dijet resonances in
  proton-proton collisions at $\sqrt{s}$ = 13 TeV and constraints on dark
  matter and other models}'',} \textit{ Phys. Lett. B} \textbf{ 769} (2017)
  520,
  \href{http://dx.doi.org/10.1016/j.physletb.2017.02.012}{\doi{10.1016/j.physletb.2017.02.012}},
\href{http://www.arXiv.org/abs/1611.03568}{\texttt{arXiv:1611.03568}}.
%%CITATION = ARXIV:1611.03568;%%.

\bibitem{Khachatryan:2015dcf}
\hrefCMSnoop {}{{CMS Collaboration}, ``Search for narrow resonances decaying to
  dijets in proton-proton collisions at {$\sqrt{s} =$ 13\TeV}'',} \textit{
  Phys. Rev. Lett.} \textbf{ 116} (2016) 071801,
  \href{http://dx.doi.org/10.1103/PhysRevLett.116.071801}{\doi{10.1103/PhysRevLett.116.071801}},
\href{http://www.arXiv.org/abs/1512.01224}{\texttt{arXiv:1512.01224}}.
%%CITATION = ARXIV:1512.01224;%%.

\bibitem{ATLAS:2015nsi}
\hrefCMSnoop {}{{ATLAS Collaboration}, ``{Search for new phenomena in dijet
  mass and angular distributions from pp collisions at $\sqrt{s}=$ 13 TeV with
  the ATLAS detector}'',} \textit{ Phys. Lett. B} \textbf{ 754} (2016) 302,
  \href{http://dx.doi.org/10.1016/j.physletb.2016.01.032}{\doi{10.1016/j.physletb.2016.01.032}},
\href{http://www.arXiv.org/abs/1512.01530}{\texttt{arXiv:1512.01530}}.
%%CITATION = ARXIV:1512.01530;%%.

\bibitem{Khachatryan:2016ecr}
\hrefCMSnoop {}{{CMS Collaboration}, ``Search for narrow resonances in dijet
  final states at {$\sqrt{s}=$ 8\TeV} with the novel {CMS} technique of data
  scouting'',} \textit{ Phys. Rev. Lett.} \textbf{ 117} (2016) 031802,
  \href{http://dx.doi.org/10.1103/PhysRevLett.117.031802}{\doi{10.1103/PhysRevLett.117.031802}},
\href{http://www.arXiv.org/abs/1604.08907}{\texttt{arXiv:1604.08907}}.
%%CITATION = ARXIV:1604.08907;%%.

\bibitem{Khachatryan:2015sja}
\hrefCMSnoop {}{{CMS Collaboration}, ``{Search for resonances and quantum black
  holes using dijet mass spectra in proton-proton collisions at $\sqrt{s} = 8$
  TeV}'',} \textit{ Phys. Rev. D} \textbf{ 91} (2015) 052009,
  \href{http://dx.doi.org/10.1103/PhysRevD.91.052009}{\doi{10.1103/PhysRevD.91.052009}},
\href{http://www.arXiv.org/abs/1501.04198}{\texttt{arXiv:1501.04198}}.
%%CITATION = ARXIV:1501.04198;%%.

\bibitem{Aad:2014aqa}
\hrefCMSnoop {}{{ATLAS Collaboration}, ``{Search for new phenomena in the dijet
  mass distribution using pp collision data at $\sqrt{s}=8$ TeV with the ATLAS
  detector}'',} \textit{ Phys. Rev. D} \textbf{ 91} (2015) 052007,
  \href{http://dx.doi.org/10.1103/PhysRevD.91.052007}{\doi{10.1103/PhysRevD.91.052007}},
\href{http://www.arXiv.org/abs/1407.1376}{\texttt{arXiv:1407.1376}}.
%%CITATION = ARXIV:1407.1376;%%.

\bibitem{Chatrchyan:2013qhXX}
\hrefCMSnoop {}{{CMS Collaboration}, ``{Search for narrow resonances using the
  dijet mass spectrum in pp collisions at $\sqrt{s}$ = 8 TeV}'',} \textit{
  Phys. Rev. D} \textbf{ 87} (2013) 114015,
  \href{http://dx.doi.org/10.1103/PhysRevD.87.114015}{\doi{10.1103/PhysRevD.87.114015}},
  \href{http://www.arXiv.org/abs/1302.4794}{\texttt{arXiv:1302.4794}}.

\bibitem{CMS:2012yf}
\hrefCMSnoop {}{{CMS Collaboration}, ``{Search for narrow resonances and
  quantum black holes in inclusive and b-tagged dijet mass spectra from pp
  collisions at $\sqrt{s}=7$ TeV}'',} \textit{ JHEP} \textbf{ 01} (2013) 013,
  \href{http://dx.doi.org/10.1007/JHEP01(2013)013}{\doi{10.1007/JHEP01(2013)013}},
  \href{http://www.arXiv.org/abs/1210.2387}{\texttt{arXiv:1210.2387}}.

\bibitem{Aad201237}
\hrefCMSnoop {}{{ATLAS Collaboration}, ``Search for new physics in the dijet
  mass distribution using 1 fb$^{-1}$ of $pp$ collision data at $\sqrt{s}$ = 7
  {TeV} collected by the {ATLAS} detector'',} \textit{ Phys. Lett. B} \textbf{
  708} (2012) 37,
  \href{http://dx.doi.org/10.1016/j.physletb.2012.01.035}{\doi{10.1016/j.physletb.2012.01.035}},
\href{http://www.arXiv.org/abs/1108.6311}{\texttt{arXiv:1108.6311}}.
%%CITATION = ARXIV:1108.6311;%%.

\bibitem{ATLAS:2012pu}
\hrefCMSnoop {}{{ATLAS Collaboration}, ``{ATLAS search for new phenomena in
  dijet mass and angular distributions using pp collisions at $\sqrt{s}=7$
  TeV}'',} \textit{ JHEP} \textbf{ 01} (2013) 029,
  \href{http://dx.doi.org/10.1007/JHEP01(2013)029}{\doi{10.1007/JHEP01(2013)029}},
  \href{http://www.arXiv.org/abs/1210.1718}{\texttt{arXiv:1210.1718}}.

\bibitem{Chatrchyan2011123}
\hrefCMSnoop {}{{CMS Collaboration}, ``{Search for resonances in the dijet mass
  spectrum from 7 TeV pp collisions at CMS}'',} \textit{ Phys. Lett. B}
  \textbf{ 704} (2011) 123,
  \href{http://dx.doi.org/10.1016/j.physletb.2011.09.015}{\doi{10.1016/j.physletb.2011.09.015}},
\href{http://www.arXiv.org/abs/1107.4771}{\texttt{arXiv:1107.4771}}.
%%CITATION = ARXIV:1107.4771;%%.

\bibitem{Aad:2011aj}
\hrefCMSnoop {}{{ATLAS Collaboration}, ``{Search for new physics in dijet mass
  and angular distributions in $pp$ collisions at $\sqrt{s} = 7$ TeV measured
  with the ATLAS detector}'',} \textit{ New J. Phys.} \textbf{ 13} (2011)
  053044,
  \href{http://dx.doi.org/10.1088/1367-2630/13/5/053044}{\doi{10.1088/1367-2630/13/5/053044}},
\href{http://www.arXiv.org/abs/1103.3864}{\texttt{arXiv:1103.3864}}.
%%CITATION = ARXIV:1103.3864;%%.

\bibitem{Khachatryan:2010jd}
\hrefCMSnoop {}{{CMS Collaboration}, ``Search for dijet resonances in 7 {TeV}
  pp collisions at {CMS}'',} \textit{ Phys. Rev. Lett.} \textbf{ 105} (2010)
  211801,
  \href{http://dx.doi.org/10.1103/PhysRevLett.105.211801}{\doi{10.1103/PhysRevLett.105.211801}},
  \href{http://www.arXiv.org/abs/1010.0203}{\texttt{arXiv:1010.0203}}.
[Erratum \DOI{10.1103/PhysRevLett.106.029902}].
%%CITATION = ARXIV:1010.0203;%%.

\bibitem{ATLAS2010}
\hrefCMSnoop {}{{ATLAS Collaboration}, ``Search for new particles in two-jet
  final states in 7 {TeV} proton-proton collisions with the {ATLAS} detector at
  the {LHC}'',} \textit{ Phys. Rev. Lett.} \textbf{ 105} (2010) 161801,
  \href{http://dx.doi.org/10.1103/PhysRevLett.105.161801}{\doi{10.1103/PhysRevLett.105.161801}},
\href{http://www.arXiv.org/abs/1008.2461}{\texttt{arXiv:1008.2461}}.
%%CITATION = ARXIV:1008.2461;%%.

\bibitem{Harris:2011bh}
\hrefCMSnoop {}{R.~M. Harris and K.~Kousouris, ``Searches for dijet resonances
  at hadron colliders'',} \textit{ Int. J. Mod. Phys. A} \textbf{ 26} (2011)
  5005,
  \href{http://dx.doi.org/10.1142/S0217751X11054905}{\doi{10.1142/S0217751X11054905}},
\href{http://www.arXiv.org/abs/1110.5302}{\texttt{arXiv:1110.5302}}.
%%CITATION = 1110.5302;%%.

\bibitem{Chivukula:2014pma}
\hrefCMSnoop {}{R.~S. Chivukula, E.~H. Simmons, and N.~Vignaroli,
  ``{Distinguishing dijet resonances at the LHC}'',} \textit{ Phys. Rev. D}
  \textbf{ 91} (2015) 055019,
  \href{http://dx.doi.org/10.1103/PhysRevD.91.055019}{\doi{10.1103/PhysRevD.91.055019}},
\href{http://www.arXiv.org/abs/1412.3094}{\texttt{arXiv:1412.3094}}.
%%CITATION = ARXIV:1412.3094;%%.

\bibitem{refCMS}
\hrefCMSnoop {}{{CMS Collaboration}, ``The {CMS} experiment at the {CERN}
  {LHC}'',} \textit{ JINST} \textbf{ 3} (2008) S08004,
\href{http://dx.doi.org/10.1088/1748-0221/3/08/S08004}{\doi{10.1088/1748-0221/3/08/S08004}}.
%%CITATION = JINST,3,S08004;%%.

\bibitem{CMS-PRF-14-001}
\hrefCMSnoop {}{{CMS Collaboration}, ``{Particle-flow reconstruction and global
  event description with the CMS detector}'',} \textit{ JINST} \textbf{ 12}
  (2017) P10003,
  \href{http://dx.doi.org/10.1088/1748-0221/12/10/P10003}{\doi{10.1088/1748-0221/12/10/P10003}},
\href{http://www.arXiv.org/abs/1706.04965}{\texttt{arXiv:1706.04965}}.
%%CITATION = ARXIV:1706.04965;%%.

\bibitem{Cacciari:2005hq}
\hrefCMSnoop {}{M.~Cacciari and G.~P. Salam, ``Dispelling the {$N^{3}$} myth
  for the $k_\mathrm{t}$ jet-finder'',} \textit{ Phys. Lett. B} \textbf{ 641}
  (2006) 57,
  \href{http://dx.doi.org/10.1016/j.physletb.2006.08.037}{\doi{10.1016/j.physletb.2006.08.037}},
\href{http://www.arXiv.org/abs/hep-ph/0512210}{\texttt{arXiv:hep-ph/0512210}}.
%%CITATION = HEP-PH/0512210;%%.

\bibitem{Cacciari:2008gp}
\hrefCMSnoop {}{M.~Cacciari, G.~P. Salam, and G.~Soyez, ``{The anti-\kt jet
  clustering algorithm}'',} \textit{ JHEP} \textbf{ 04} (2008) 063,
  \href{http://dx.doi.org/10.1088/1126-6708/2008/04/063}{\doi{10.1088/1126-6708/2008/04/063}},
  \href{http://www.arXiv.org/abs/0802.1189}{\texttt{arXiv:0802.1189}}.

\bibitem{Cacciari:2011ma}
\hrefCMSnoop {}{M.~Cacciari, G.~P. Salam, and G.~Soyez, ``{FastJet user
  manual}'',} \textit{ Eur. Phys. J. C} \textbf{ 72} (2012) 1896,
  \href{http://dx.doi.org/10.1140/epjc/s10052-012-1896-2}{\doi{10.1140/epjc/s10052-012-1896-2}},
\href{http://www.arXiv.org/abs/1111.6097}{\texttt{arXiv:1111.6097}}.
%%CITATION = ARXIV:1111.6097;%%.

\bibitem{jetarea_fastjet_pu}
\hrefCMSnoop {}{M.~Cacciari and G.~P. Salam, ``Pileup subtraction using jet
  areas'',} \textit{ Phys. Lett. B} \textbf{ 659} (2008) 119,
  \href{http://dx.doi.org/10.1016/j.physletb.2007.09.077}{\doi{10.1016/j.physletb.2007.09.077}},
\href{http://www.arXiv.org/abs/0707.1378}{\texttt{arXiv:0707.1378}}.
%%CITATION = ARXIV:0707.1378;%%.

\bibitem{Khachatryan:2016kdb}
\hrefCMSnoop {}{{CMS Collaboration}, ``{Jet energy scale and resolution in the
  CMS experiment in pp collisions at 8 TeV}'',} \textit{ JINST} \textbf{ 12}
  (2017) P02014,
  \href{http://dx.doi.org/10.1088/1748-0221/12/02/P02014}{\doi{10.1088/1748-0221/12/02/P02014}},
\href{http://www.arXiv.org/abs/1607.03663}{\texttt{arXiv:1607.03663}}.
%%CITATION = ARXIV:1607.03663;%%.

\bibitem{Khachatryan:2016bia}
\hrefCMSnoop {}{{CMS Collaboration}, ``{The CMS trigger system}'',} \textit{
  JINST} \textbf{ 12} (2017) P01020,
  \href{http://dx.doi.org/10.1088/1748-0221/12/01/P01020}{\doi{10.1088/1748-0221/12/01/P01020}},
\href{http://www.arXiv.org/abs/1609.02366}{\texttt{arXiv:1609.02366}}.
%%CITATION = ARXIV:1609.02366;%%.

\bibitem{CMS:2017wyc}
\href {https://cds.cern.ch/record/2256875}{{CMS Collaboration}, ``{Jet
  algorithms performance in 13 TeV data}'',} CMS Physics Analysis Summary
  CMS-PAS-JME-16-003, CERN, Geneva, 2017.

\bibitem{Sjostrand:2014zea}
T.~Sj{\"o}strand\hrefCMSnoop {}{ {et~al.}, ``{An introduction to PYTHIA
  8.2}'',} \textit{ Comput. Phys. Commun.} \textbf{ 191} (2015) 159,
  \href{http://dx.doi.org/10.1016/j.cpc.2015.01.024}{\doi{10.1016/j.cpc.2015.01.024}},
\href{http://www.arXiv.org/abs/1410.3012}{\texttt{arXiv:1410.3012}}.
%%CITATION = ARXIV:1410.3012;%%.

\bibitem{Khachatryan:2015pea}
\hrefCMSnoop {}{{CMS Collaboration}, ``{Event generator tunes obtained from
  underlying event and multiparton scattering measurements}'',} \textit{ Eur.
  Phys. J. C} \textbf{ 76} (2016) 155,
  \href{http://dx.doi.org/10.1140/epjc/s10052-016-3988-x}{\doi{10.1140/epjc/s10052-016-3988-x}},
\href{http://www.arXiv.org/abs/1512.00815}{\texttt{arXiv:1512.00815}}.
%%CITATION = ARXIV:1512.00815;%%.

\bibitem{Skands:2014pea}
\hrefCMSnoop {}{P.~Skands, S.~Carrazza, and J.~Rojo, ``{Tuning PYTHIA 8.1: the
  Monash 2013 tune}'',} \textit{ Eur. Phys. J. C} \textbf{ 74} (2014) 3024,
  \href{http://dx.doi.org/10.1140/epjc/s10052-014-3024-y}{\doi{10.1140/epjc/s10052-014-3024-y}},
  \href{http://www.arXiv.org/abs/1404.5630}{\texttt{arXiv:1404.5630}}.

\bibitem{Ball:2012cx}
\hrefCMSnoop {}{{NNPDF} Collaboration, ``{Parton distributions with LHC
  data}'',} \textit{ Nucl. Phys. B} \textbf{ 867} (2013) 244,
  \href{http://dx.doi.org/10.1016/j.nuclphysb.2012.10.003}{\doi{10.1016/j.nuclphysb.2012.10.003}},
\href{http://www.arXiv.org/abs/1207.1303}{\texttt{arXiv:1207.1303}}.
%%CITATION = ARXIV:1207.1303;%%.

\bibitem{refGEANT}
\hrefCMSnoop {}{{GEANT4} Collaboration, ``{GEANT4} --- a simulation toolkit'',}
  \textit{ Nucl. Instrum. Meth. A} \textbf{ 506} (2003) 250,
\href{http://dx.doi.org/10.1016/S0168-9002(03)01368-8}{\doi{10.1016/S0168-9002(03)01368-8}}.
%%CITATION = NUIMA,A506,250;%%.

\bibitem{ref_dobrescu}
\hrefCMSnoop {}{B.~A. Dobrescu, R.~M. Harris, and J.~Isaacson, ``{Ultraheavy
  resonances at the LHC: beyond the QCD background}'',} (2018).
\href{http://www.arXiv.org/abs/1810.09429}{\texttt{arXiv:1810.09429}}.
%%CITATION = ARXIV:1810.09429;%%.

\bibitem{refCDFrun2}
\hrefCMSnoop {}{{CDF} Collaboration, ``Search for new particles decaying into
  dijets in proton-antiproton collisions at $\sqrt{s} = 1.96$~{TeV}'',}
  \textit{ Phys. Rev. D} \textbf{ 79} (2009) 112002,
  \href{http://dx.doi.org/10.1103/PhysRevD.79.112002}{\doi{10.1103/PhysRevD.79.112002}},
\href{http://www.arXiv.org/abs/0812.4036}{\texttt{arXiv:0812.4036}}.
%%CITATION = ARXIV:0812.4036;%%.

\bibitem{Nason:2004rx}
\hrefCMSnoop {}{P.~Nason, ``{A New method for combining NLO QCD with shower
  Monte Carlo algorithms}'',} \textit{ JHEP} \textbf{ 11} (2004) 040,
  \href{http://dx.doi.org/10.1088/1126-6708/2004/11/040}{\doi{10.1088/1126-6708/2004/11/040}},
\href{http://www.arXiv.org/abs/hep-ph/0409146}{\texttt{arXiv:hep-ph/0409146}}.
%%CITATION = HEP-PH/0409146;%%.

\bibitem{Frixione:2007vw}
\hrefCMSnoop {}{S.~Frixione, P.~Nason, and C.~Oleari, ``{Matching NLO QCD
  computations with parton shower simulations: the POWHEG method}'',} \textit{
  JHEP} \textbf{ 11} (2007) 070,
  \href{http://dx.doi.org/10.1088/1126-6708/2007/11/070}{\doi{10.1088/1126-6708/2007/11/070}},
\href{http://www.arXiv.org/abs/0709.2092}{\texttt{arXiv:0709.2092}}.
%%CITATION = ARXIV:0709.2092;%%.

\bibitem{Alioli:2010xd}
\hrefCMSnoop {}{S.~Alioli, P.~Nason, C.~Oleari, and E.~Re, ``{A general
  framework for implementing NLO calculations in shower Monte Carlo programs:
  the POWHEG BOX}'',} \textit{ JHEP} \textbf{ 06} (2010) 043,
  \href{http://dx.doi.org/10.1007/JHEP06(2010)043}{\doi{10.1007/JHEP06(2010)043}},
\href{http://www.arXiv.org/abs/1002.2581}{\texttt{arXiv:1002.2581}}.
%%CITATION = ARXIV:1002.2581;%%.

\bibitem{Dittmaier:2012kx}
\hrefCMSnoop {}{S.~Dittmaier, A.~Huss, and C.~Speckner, ``{Weak radiative
  corrections to dijet production at hadron colliders}'',} \textit{ JHEP}
  \textbf{ 11} (2012) 095,
  \href{http://dx.doi.org/10.1007/JHEP11(2012)095}{\doi{10.1007/JHEP11(2012)095}},
\href{http://www.arXiv.org/abs/1210.0438}{\texttt{arXiv:1210.0438}}.
%%CITATION = ARXIV:1210.0438;%%.

\bibitem{CMS-PAS-LUM-17-001}
\href {https://cds.cern.ch/record/2257069}{{CMS Collaboration}, ``{CMS}
  luminosity measurements for the 2016 data taking period'',} CMS Physics
  Analysis Summary CMS-PAS-LUM-17-001, CERN, Geneva, 2017.

\bibitem{CMS-PAS-LUM-18-002}
\href {https://cds.cern.ch/record/2676164}{{CMS Collaboration}, ``{CMS}
  luminosity measurements for the 2018 data taking period'',} CMS Physics
  Analysis Summary CMS-PAS-LUM-18-002, CERN, Geneva, 2017.

\bibitem{CMS-PAS-LUM-17-004}
\href {https://cds.cern.ch/record/2621960}{{CMS Collaboration}, ``{CMS}
  luminosity measurements for the 2017 data taking period'',} CMS Physics
  Analysis Summary CMS-PAS-LUM-17-004, CERN, Geneva, 2017.

\bibitem{Junk1999}
\hrefCMSnoop {}{T.~Junk, ``{Confidence level computation for combining searches
  with small statistics}'',} \textit{ Nucl. Instr. Meth. A} \textbf{ 434}
  (1999) 435,
  \href{http://dx.doi.org/10.1016/S0168-9002(99)00498-2}{\doi{10.1016/S0168-9002(99)00498-2}},
\href{http://www.arXiv.org/abs/hep-ex/9902006}{\texttt{arXiv:hep-ex/9902006}}.
%%CITATION = HEP-EX/9902006;%%.

\bibitem{bib-cls}
\hrefCMSnoop {}{A.~L. Read, ``{Presentation of search results: the {$CL_s$}
  technique}'',} \textit{ J. Phys. G} \textbf{ 28} (2002) 2693,
\href{http://dx.doi.org/10.1088/0954-3899/28/10/313}{\doi{10.1088/0954-3899/28/10/313}}.
%%CITATION = JPHGB,G28,2693;%%.

\bibitem{ATLAS:1379837}
\href {https://cds.cern.ch/record/1379837}{{LHC Higgs Combination Group},
  ``{Procedure for the LHC Higgs boson search combination in Summer 2011}'',}
  Technical Report CMS-NOTE-2011-005, ATL-PHYS-PUB-2011-11, CERN, Geneva, 2011.

\bibitem{Cowan:2010js}
\hrefCMSnoop {}{G.~Cowan, K.~Cranmer, E.~Gross, and O.~Vitells, ``Asymptotic
  formulae for likelihood-based tests of new physics'',} \textit{ Eur. Phys. J.
  C} \textbf{ 71} (2011) 1554,
  \href{http://dx.doi.org/10.1140/epjc/s10052-011-1554-0}{\doi{10.1140/epjc/s10052-011-1554-0}},
  \href{http://www.arXiv.org/abs/1007.1727}{\texttt{arXiv:1007.1727}}.
[Erratum: \DOI{10.1140/epjc/s10052-013-2501-z}].
%%CITATION = ARXIV:1007.1727;%%.

\bibitem{refCTEQ}
J.~Pumplin\hrefCMSnoop {}{ {et~al.}, ``{New generation of parton distributions
  with uncertainties from global QCD analysis}'',} \textit{ JHEP} \textbf{ 07}
  (2002) 012,
  \href{http://dx.doi.org/10.1088/1126-6708/2002/07/012}{\doi{10.1088/1126-6708/2002/07/012}},
  \href{http://www.arXiv.org/abs/hep-ph/0201195}{\texttt{arXiv:hep-ph/0201195}}.

\bibitem{Barger:1987nn}
V.~D. Barger and R.~J.~N. Phillips, ``Collider physics, updated edition,
  frontiers in physics volume 71''.
\newblock Westview Press, Boulder, Colorado, 1996.
\newblock
ISBN~0-201-14945-1,
%%CITATION = INSPIRE-255673;%%.

\bibitem{Alwall:2014hca}
J.~Alwall\hrefCMSnoop {}{ {et~al.}, ``{The automated computation of tree-level
  and next-to-leading order differential cross sections, and their matching to
  parton shower simulations}'',} \textit{ JHEP} \textbf{ 07} (2014) 079,
  \href{http://dx.doi.org/10.1007/JHEP07(2014)079}{\doi{10.1007/JHEP07(2014)079}},
\href{http://www.arXiv.org/abs/1405.0301}{\texttt{arXiv:1405.0301}}.
%%CITATION = ARXIV:1405.0301;%%.

\bibitem{Dobrescu:2013coa}
\hrefCMSnoop {}{B.~A. Dobrescu and F.~Yu, ``{Coupling-mass mapping of dijet
  peak searches}'',} \textit{ Phys. Rev. D} \textbf{ 88} (2013) 035021,
  \href{http://dx.doi.org/10.1103/PhysRevD.88.035021}{\doi{10.1103/PhysRevD.88.035021}},
  \href{http://www.arXiv.org/abs/1306.2629}{\texttt{arXiv:1306.2629}}.
[Erratum: \DOI{10.1103/PhysRevD.90.079901}].
%%CITATION = ARXIV:1306.2629;%%.

\end{thebibliography}\endgroup

\cleardoublepage \appendix\section{The CMS Collaboration \label{app:collab}}\begin{sloppypar}\hyphenpenalty=5000\widowpenalty=500\clubpenalty=5000\vskip\cmsinstskip
\textbf{Yerevan Physics Institute, Yerevan, Armenia}\\*[0pt]
A.M.~Sirunyan$^{\textrm{\dag}}$, A.~Tumasyan
\vskip\cmsinstskip
\textbf{Institut f\"{u}r Hochenergiephysik, Wien, Austria}\\*[0pt]
W.~Adam, F.~Ambrogi, T.~Bergauer, M.~Dragicevic, J.~Er\"{o}, A.~Escalante~Del~Valle, M.~Flechl, R.~Fr\"{u}hwirth\cmsAuthorMark{1}, M.~Jeitler\cmsAuthorMark{1}, N.~Krammer, I.~Kr\"{a}tschmer, D.~Liko, T.~Madlener, I.~Mikulec, N.~Rad, J.~Schieck\cmsAuthorMark{1}, R.~Sch\"{o}fbeck, M.~Spanring, D.~Spitzbart, W.~Waltenberger, C.-E.~Wulz\cmsAuthorMark{1}, M.~Zarucki
% ...
\vskip\cmsinstskip
\dag: Deceased\\
1:  Also at Vienna University of Technology, Vienna, Austria\\
% ...
\end{sloppypar}
%%% END EDITABLE REGION %%%
\end{document}